\documentclass[conference]{IEEEtran}
\IEEEoverridecommandlockouts
\usepackage{cite}
\usepackage{amsmath,amssymb,amsfonts}
\usepackage{graphicx}
\usepackage{textcomp}
\usepackage{xcolor}
\def\BibTeX{{\rm B\kern-.05em{\sc i\kern-.025em b}\kern-.08em
    T\kern-.1667em\lower.7ex\hbox{E}\kern-.125emX}}

\usepackage{hyperref}
\usepackage[utf8]{inputenc}
\usepackage[T1]{fontenc}
\usepackage{graphicx} 
\usepackage{amsmath}
\usepackage{algorithm}            
\usepackage[noend]{algpseudocode}        
\usepackage{algorithmicx}         
\algdef{SE}[DOWHILE]{Do}{doWhile}{\algorithmicdo}[1]{\algorithmicwhile\ #1}
\algrenewcommand\algorithmicrequire{\textbf{Input:}}
\algrenewcommand\algorithmicensure{\textbf{Output:}}

\usepackage{svg}
\usepackage{float}
\usepackage{tikz}
\usepackage{afterpage}
\usepackage{xcolor} 
\usepackage{soul}

\usepackage{orcidlink}
\usepackage{stfloats}

\begin{document}

\title{A High-Performance Training-Free Pipeline for Robust Random Telegraph Signal Characterization via Adaptive Wavelet-Based Denoising and Bayesian Digitization Methods\\
}

\author{
\IEEEauthorblockN{
Tonghe Bai\textsuperscript{1,2}\,\orcidlink{0009-0003-7338-2986},
Ayush Kapoor\textsuperscript{2},
and Na Young Kim\textsuperscript{1,2,3}\,\orcidlink{0000-0002-0290-7728}
}
\IEEEauthorblockA{
\textsuperscript{1}\textit{Institute for Quantum Computing, University of Waterloo, Waterloo, ON, N2L 3G1, Canada}\\
\textsuperscript{2}\textit{Department of Electrical and Computer Engineering, University of Waterloo, Waterloo, ON, N2L 3G1, Canada}\\
\textsuperscript{3}\textit{Waterloo Institute for Nanotechnology, University of Waterloo, Waterloo, ON, N2L 3G1, Canada}\\
Emails: t4bai@uwaterloo.ca, a22kapoo@uwaterloo.ca, nayoung.kim@uwaterloo.ca
}
}

\maketitle

\begin{abstract}
Random telegraph signal (RTS) analysis is increasingly important for characterizing meaningful temporal fluctuations in physical, chemical, and biological systems. The simplest RTS arises from discrete stochastic switching events between two binary states, quantified by their transition amplitude and dwell times in each state. Quantitative analysis of RTSs provides valuable insights into microscopic processes such as charge trapping in semiconductors. However, analyzing RTS becomes considerably complex when signals exhibit multi-level structures or are corrupted by background white or pink noise. To address these challenges and support high-throughput RTS characterization, we propose a modular, training-free signal processing pipeline that integrates adaptive dual-tree complex wavelet transform (DTCWT) denoising with a lightweight Bayesian digitization strategy. The adaptive DTCWT denoiser incorporates autonomous parameter selection rules for its decomposition level and thresholds, optimizing white noise suppression without manual tuning. Complementing this stage, our Bayesian digitizer formulates RTS level assignment as a probabilistic latent-state inference problem incorporating temporal regularization without iterative optimization, effectively resolving binary trap states even under residual notorious background pink noise. Quantitative benchmarking on large synthetic datasets with known ground truth demonstrates improved RTS reconstruction accuracy, trap-state resolution, and dwell-time estimation across diverse noise regimes and multi-trap scenarios, while achieving up to 83$\times$ speedups over classical and neural baselines. Qualitative validation on experimental RTS data when no ground truth is available illustrates practical usability and flexibility for real-time or large-scale analysis in real measurement settings. Together, the proposed framework establishes a scalable and reproducible foundation for autonomous RTS analysis and systematic benchmarking, with potential to support future extensions toward more complex and device-specific RTS studies.
\end{abstract}


\begin{figure*}[!]
    \centering
    \includegraphics[width=0.8\linewidth]{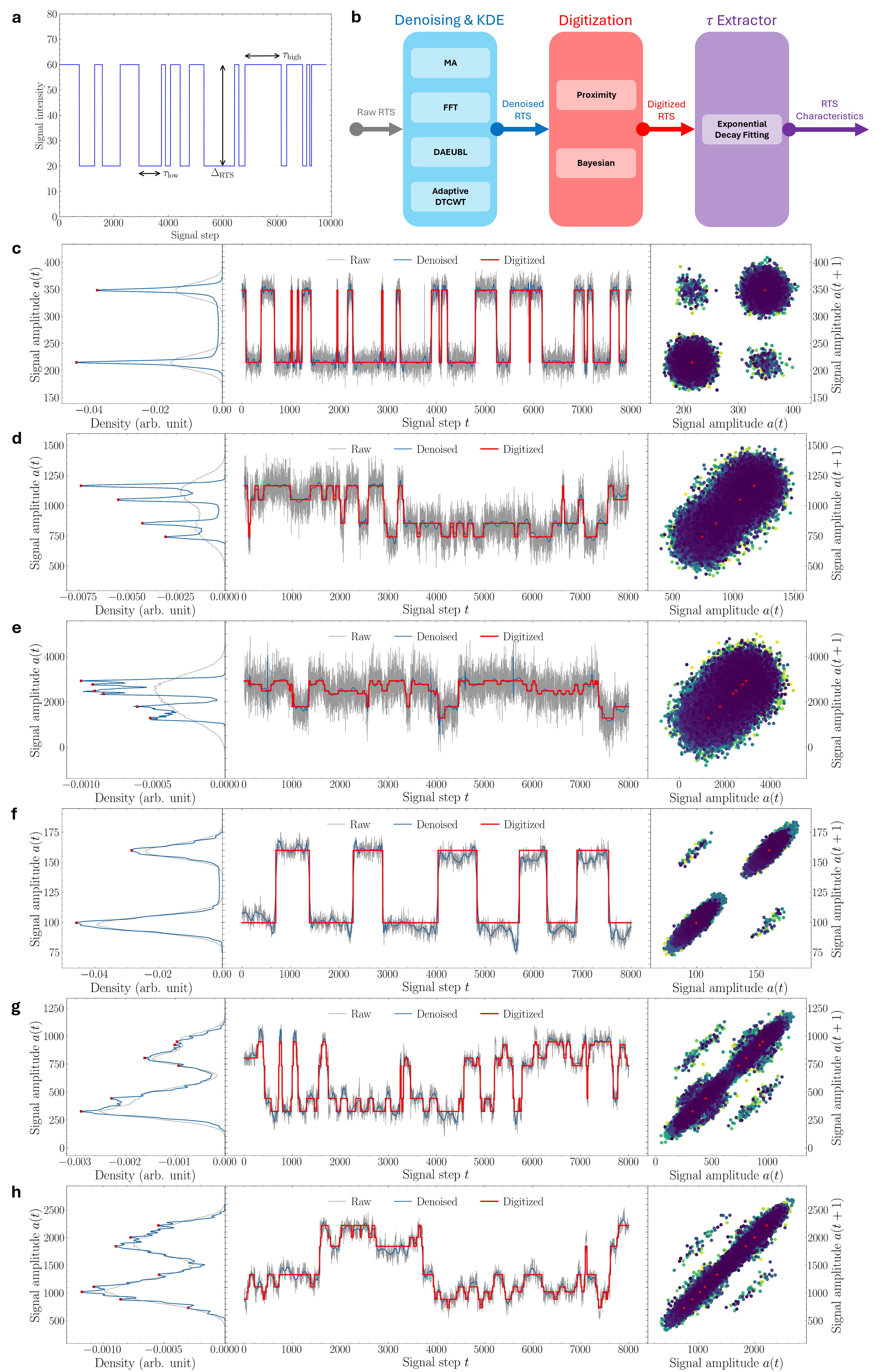}
    \caption{\textbf{(a)} A noiseless simple 2-level RTS with the definition of $\Delta_{\text{RTS}}$, ${\tau}_{\text{high}}$, and ${\tau}_{\text{low}}$. Examples of processed synthesized RTS with noisy RTS (grey), denoised RTS (blue), and digitized RTS by our \textbf{DTCWT + Bayesian} method (red) on middle subplot, cropped for better visualization; kernel density estimation (KDE) plot on left subplot; time-lag plot on right subplot for the entire RTS duration. \textbf{(b)} The workflow of our three-stage RTS analysis pipeline. \textbf{(c)} 1-trap RTS with $\eta_{\text{wn}}=10\%$. \textbf{(d)} 2-trap RTS with $\eta_{\text{wn}}=20\%$. \textbf{(e)} 3-trap RTS with $\eta_{\text{wn}}=30\%$. \textbf{(f)} 1-trap RTS with $\eta_{\text{pn}}=10\%$. \textbf{(g)} 2-trap RTS with $\eta_{\text{pn}}=10\%$. \textbf{(h)} 3-trap RTS with $\eta_{\text{pn}}=10\%$.}
    \label{fig:fig1_rts_examples}
\end{figure*}

\section{Introduction}

Random telegraph signals (RTS) are time-dependent signals that exhibit discrete, stochastic switching events between two distinct levels. Each transition is characterized by a specific amplitude, denoted as $\Delta_{\text{RTS}}$, and by dwell times $\tau_{\text{low}}$ and $\tau_{\text{high}}$, which quantify how long the signal resides in the low and high levels, respectively (\autoref{fig:fig1_rts_examples}a). Such fluctuations have been reported across a wide range of systems, from ionic transport in electrochemical settings to active biomolecular processes \cite{biochem} and stochastic gene expression in biology \cite{gene}, as well as signal instabilities in diodes \cite{apd}, sensors \cite{sensors}, and even quantum bit (qubit) readouts in quantum computing \cite{rts.SR.ref3, rts.SR.ref5}. In the realm of electronics, RTS often arises from single-charge trapping and detrapping random processes and is observed in atomic-scale devices such as nanoscale metal-oxide-semiconductor field-effect transistors \cite{transistor, mosfets, rts.SR.ref1} and complementary metal-oxide-semiconductor circuits for image sensors \cite{rts.SR.ref2}. In modern semiconductor research, this phenomenon is particularly critical, as devices shrink to the nanoscale and become increasingly susceptible to single-charge events and defect-induced noise \cite{rts_noise}. Across these diverse contexts, the discrete fluctuations originate from fundamental microscopic dynamics, and understanding them is essential for both device design and theoretical insight.

The goal of RTS analysis is to extract the transition amplitude ($\Delta_{\text{RTS}}$) and the exponential dwell-time distributions, which reflect the memoryless nature of a Poissonian process. This is essential for identifying the physical variables that govern RTS and for developing reliable models that capture the underlying microscopic dynamics of the nanoscale systems. Yet, analyzing and modeling RTS processes poses unique challenges when other noise sources are present at comparable or higher amplitudes, as the RTS fluctuations can be masked or exacerbated under such high-noise conditions. Among the various noise sources in electronic measurements, white noise and pink noise are especially unavoidable. White noise has a flat spectral density, with power evenly distributed across all frequencies, whereas pink noise exhibits a $1/f$-spectrum concentrating more power at lower frequencies \cite{low_freq}. This low-frequency dominance makes pink noise particularly problematic for RTS analysis, as it masks slow transitions and obscures the identification of individual RTS levels and switching events. To address these challenges, we aim to develop an efficient and systematic RTS analysis protocol capable of reliably handling multi-level RTSs in the presence of both white and pink noise.


As RTS signals grow more complex and involve multiple levels, sophisticated analysis protocols are in demand beyond generic signal processing methods, with a particular emphasis on denoising and digitization. Traditional denoising techniques include moving average (MA) filtering \cite{MA.reram}, fast Fourier transform (FFT) with frequency filtering, empirical mode decomposition \cite{emd}, and various wavelet-based methods \cite{wavelet_methods}. These techniques provide baseline denoising capabilities, especially effective for signals corrupted by white noise. Wavelet-based approaches have also been widely employed in other signal denoising and multiscale analysis applications involving discrete and multilevel data, including large-scale N-body and cosmological simulations, where they have been employed to mitigate discreteness effects and analyze scale-dependent structure \cite{Alessandro_Romeo_1,Alessandro_Romeo_2,Alessandro_Romeo_3}. Probabilistic models, such as hidden Markov models (HMMs) \cite{hmm.nature, hmm.harvard, hmm.language} and Gaussian mixture models (GMMs) \cite{rts.SR}, have also been applied to RTS analysis, offering strong interpretability of RTS and quantification of RTS parameters. Unfortunately, these approaches often require careful parameter initialization, limiting their scalability and robustness in complex RTS environments. When confronting pink noise and multi-trap scenarios, these classical approaches fall short.

Besides the challenges of noise masking, another important difficulty in RTS analysis is accurately assigning each discrete level to its corresponding trap state, especially in multi-trap scenarios where level overlap and noise can confound traditional digitization methods. Although considerable work has been done to quantify key RTS parameters, such as the amplitude $\Delta_{\text{RTS}}$ and the mean dwell times $\bar{\tau}_{\text{high}}$ and $\bar{\tau}_{\text{low}}$ \cite{bowen.thesis, rts.SR}, noise remains a major barrier to accuracy. For instance, density-based techniques such as kernel density estimation (KDE) can identify $\Delta_{\text{RTS}}$ through clear multimodal peaks in amplitude statistics, but only under well-denoised conditions, while accurate estimation of dwell times likewise depends on the quality of denoising and digitization.

Recently, neural network (NN)-based methods have been proposed and applied to real RTS data from nanoscale electronic devices \cite{rts_ml_RNN, rts_ml_clustering, rts_ml_SOM}. In particular, NN-based denoising methods have gained attention due to their ability to generalize across diverse noise conditions, such as multi-trap or pink noise, without requiring manual parameter tuning. Architectures such as Recurrent neural networks \cite{rts_ml_RNN} and specialized autoencoder models like DAEUBL - a denoising autoencoder based on U-Net and bidirectional long short-term memory layers \cite{bowen.thesis, rts.SR} has demonstrated improved performance in digitizing clean RTS from noisy measurements, often outperforming classical filters under complex noise conditions. Self-organizing maps, as an alternative unsupervised NN approach, have also been employed for RTS pattern recognition in resistive random-access memories \cite{rts_ml_SOM}. Beyond NNs, another machine learning (ML) technique, such as K-Medoids clustering, has been applied to real cryogenic transistor data for localizing complex RTS features \cite{rts_ml_clustering}. Meanwhile, a more general ML framework for RTS analysis is emerging as well \cite{rts_ml_RTNinja}. Despite these advances, NN-based methods still face critical drawbacks. Their inference speed and memory consumption become prohibitive when applied to long-duration or high-resolution signals. For example, a 100-second RTS recorded at 10 ns resolution yields 10 billion time steps, well beyond the processing capabilities of typical NN models like DAEUBL within a reasonable amount of time and memory, even on NVIDIA V100 GPUs.

While denoising has seen significant innovation with these advanced neural-based and wavelet-based models, RTS digitization, the process of mapping denoised signals to discrete levels and decomposing them into binary sequences for individual traps, has received comparatively less systematic attention. Especially, the task becomes particularly challenging when signals involve more than 2 levels. Current practice primarily relies on simple thresholding methods \cite{qd.SR} or proximity-based peak-matching heuristics \cite{bowen.thesis, rts.SR}. Some probabilistic-based methods have been explored in the context of RTS noise analysis \cite{rtn.bayesian, rtn.bayesian2}, offering a way to model and interpret fluctuating signals in a statistically grounded manner. Bayesian approaches, particularly suited for handling uncertain and noisy data, are valuable for extracting meaningful information from RTSs, where noise often obscures state transitions \cite{rtn.bayesian, rtn.bayesian2}, but existing formulations typically rely on complex models or require extensive tuning and prior assumptions \cite{mcmc1, mcmc2}.

Finally, a unified framework for evaluating RTS analysis performance remains missing from the literature, as prior studies typically assess only a subset of metrics, without a comprehensive or standardized evaluation protocol. To address this gap, we introduce a fully specified RTS signal processing pipeline that integrates denoising and digitization within a single modular framework, together with a comprehensive suite of quality and performance metrics. In particular, we define a 7-category benchmarking regime including various novel measures such as trap count error ($N_{\text{trap}}$ error), to thoroughly evaluate the ability of each algorithm to characterize individual RTS traps in a systematic and reproducible manner.

Here we offer a comprehensive training-free and modular RTS analysis pipeline designed for robust denoising, high-fidelity digitization, and statistical characterization. At its core are two primary algorithmic contributions: an adaptive dual-tree complex wavelet transform (DTCWT) denoiser and a Bayesian digitizer. Each contribution is selected to address limitations observed in prior methods, particularly under high-noise conditions and multi-trap scenarios. High-noise conditions, we refer to settings with a wide range of noise levels, including severe background fluctuations that can obscure step transitions—a point elaborated further in the Methods section. While wavelet-based denoising \cite{rts_wavelet} and probabilistic inference \cite{rtn.bayesian,rtn.bayesian2, mcmc1, mcmc2} have each been explored independently in prior RTS studies, our work adopts in its adaptive formulation and unified integration into an autonomous pipeline that operates without training data or learned models. Each RTS trace is processed independently, avoiding dependence on prior signal statistics and reducing the barrier for experimental adoption.

The proposed adaptive DTCWT denoiser leverages the approximate shift invariance and better transition preservation of the traditional DTCWT by incorporating automated decomposition-level selection and thresholding rules specific to RTS transitions. The benchmarking results show that these characteristics allow for efficient background noise suppression while preserving transition fidelity under a variety of noise conditions. A probabilistic model that interprets the denoised signal in terms of discrete amplitude levels found from KDE is then used to handle RTS digitization. The digitizer determines RTS levels using a Bayesian inference process that integrates data from the observed signal amplitude and prior state likelihoods, as opposed to heuristic thresholding or proximity-based peak matching. In addition to being computationally efficient and appropriate for large-scale RTS analysis, this method offers a lightweight and comprehensible substitute for more intricate generative models. \textbf{Methods} provides a detailed description of the formal probabilistic formulation and inference process.

The workflow for the RTS characterization, as illustrated in \autoref{fig:fig1_rts_examples}b, involves three primary stages:(1) denoising and KDE, (2) digitization, and (3) dwell time statistics extraction. First, a raw signal is processed using a denoising algorithm to suppress background fluctuations while preserving the discrete transitions characteristic of the RTS behavior. The denoised signal is then analyzed using statistical tools such as kernel density estimation (KDE) to identify discrete amplitude levels. In the second stage, the identified levels then serve to digitize the signal into binary sequences, each representing an individual trap. Finally, from these digitized traces, we extract key trap parameters such as their characteristic dwell times $\bar{\tau}_{\text{high}}$ and $\bar{\tau}_{\text{low}}$, corresponding to the average durations that the system stays in the low and high states in \autoref{fig:fig1_rts_examples}a. 

Reliable characterization of each trap, such as quantifying $\bar{\tau}_{\text{high}}$, $\bar{\tau}_{\text{low}}$ and $\Delta_{\text{RTS}}$ for real data is essential, as it directly informs device reliability and can guide design decisions in manufacturing. As real experimental data lack a well-defined `true value', validating our proposed pipeline through systematic performance quantification is only feasible using synthesis RTS data. \autoref{fig:fig1_rts_examples}c–h shows examples of processed synthesized RTS signals. The procedure for generating these ground-truth synthesized RTS datasets is described in \textbf{Methods}. Motivated by the need for reliable validation using controlled datasets, we next present a systematic evaluation of our proposed RTS analysis pipeline on synthesized signals spanning diverse trap numbers, noise types, and noise levels.

\section{Results}

To benchmark the pipeline, we synthesize a dataset consisting of 1,800 RTS samples, each with length $L = 100{,}000$ time steps. These samples span evenly distributed across ground truth trap counts $N_{\text{trap}} = \{1, 2, 3\}$, two background noise types (white noise (wn) and pink noise (pn)) and noise levels, $\eta_{\text{wn}}, \eta_{\text{pn}} \in [1\%, 30\%]$. Quantitative benchmarking is performed on synthetic RTS signals, as real experimental RTS data generally lack ground-truth annotations for trap number, transition amplitudes, and dwell times, making objective quantitative evaluation infeasible without extensive device-specific physical modeling. \autoref{fig:fig1_rts_examples}c–e illustrate 1-trap, 2-trap, and 3-trap RTSs ($a(t)$) injected with three levels of white noise $\eta_{\text{wn}} = 10\%, 20\%, 30\%$ respectively. As both the number of traps and noise level increase, KDE peaks in the left sub panel become less distinguishable, and the time-lag plots increasingly overlap colonies of points. For a 3-trap RTS in \autoref{fig:fig1_rts_examples}e, only 6 KDE peaks are identified, out of the 8 expected levels if three traps are independent and mutually exclusive. \autoref{fig:fig1_rts_examples}f–h correspond to 1-trap, 2-trap, and 3-trap RTSs masked by the same pink noise level, $\eta_{\text{wn}}=10\%$ in all three cases. Higher trap counts under pink noise present greater denoising challenges: residual low-frequency components can create spurious levels in the KDE. In \autoref{fig:fig1_rts_examples}g, for instance, a true 2-trap RTS may be misclassified as a 3-trap RTS due to multiple sub-peaks. Synthetic RTS signals therefore play a critical role in enabling controlled and reproducible evaluation across noise types, trap counts, and signal complexities, allowing systematic stress testing of denoising and digitization performance that is not possible with real experimental data alone. Beyond the difficulty of colored-noise masking, a further major difficulty lies in the efficient processing of very long RTS signals. In high-resolution experimental measurements, nanoscale devices can generate datasets containing millions or even billions of time steps. This scale places heavy demands on both memory and computation, making it essential to design denoising and digitization algorithms that are not only accurate but also adaptable and computationally efficient.

We emphasize that our RTS datasets are generated under more challenging conditions than prior studies \cite{bowen.thesis,rts.SR}, with a tighter definition of background noise strength $\eta_\text{wn(pn)}$ in \textbf{Methods}. Furthermore, we incorporate scaled trap amplitudes, where each additional trap has a reduced strength relative to the dominant trap, to simulate sub-dominant traps and introduce strong pink noise that can fully mask weaker transitions, thereby stressing both the denoising and digitization stages (details in \textbf{Methods}). Real experimental RTS examples are included in this work to demonstrate practical applicability of the proposed pipeline rather than to provide quantitative validation, as meaningful numerical error metrics on real RTS data typically require expert annotation and detailed device-specific physical interpretation, often forming the focus of a standalone experimental study. To demonstrate the advantages of our approach, we define two primary benchmark categories: quality and performance. These benchmarks quantitatively assess both the proposed and baseline methods in a controlled and reproducible manner, ensuring that both accuracy and computational efficiency are considered.

Four complete methods are compared across all evaluation metrics: MA denoising with proximity-based digitization (\textbf{MA + Proximity}), FFT denoising with proximity-based digitization (\textbf{FFT + Proximity}), DAEUBL denoising with proximity-based digitization (\textbf{DAEUBL + Proximity}), and our proposed DTCWT denoising combined with Bayesian digitization (\textbf{DTCWT + Bayesian}). The first two represent traditional, non-neural-network baselines. The third is a neural-network-based method—the only such method currently available for this task. While the first three combinations serve as baseline references, the fourth is our proposed approach, which is a modular and probabilistic alternative for robust RTS analysis. Our \textbf{DTCWT + Bayesian} is a coherent pipeline that pairs adaptive DTCWT denoising with a probabilistic model-based digitization scheme. It is important to clarify that digitization occurs after denoising and KDE. Thus, the digitization quality measured in our benchmarks reflects the effectiveness of the entire method pipeline, not just the digitization algorithm in isolation. Below summarizes the results of the denoising quality: \textbf{signal-to-noise ratio}, \textbf{trap number error}, and \textbf{trap transition amplitude error}; as well as the digitization quality as \textbf{trap state error}, and \textbf{mean dwell times error}.

\begin{figure*}[!] 
    \centering
    \includegraphics[width=\linewidth]{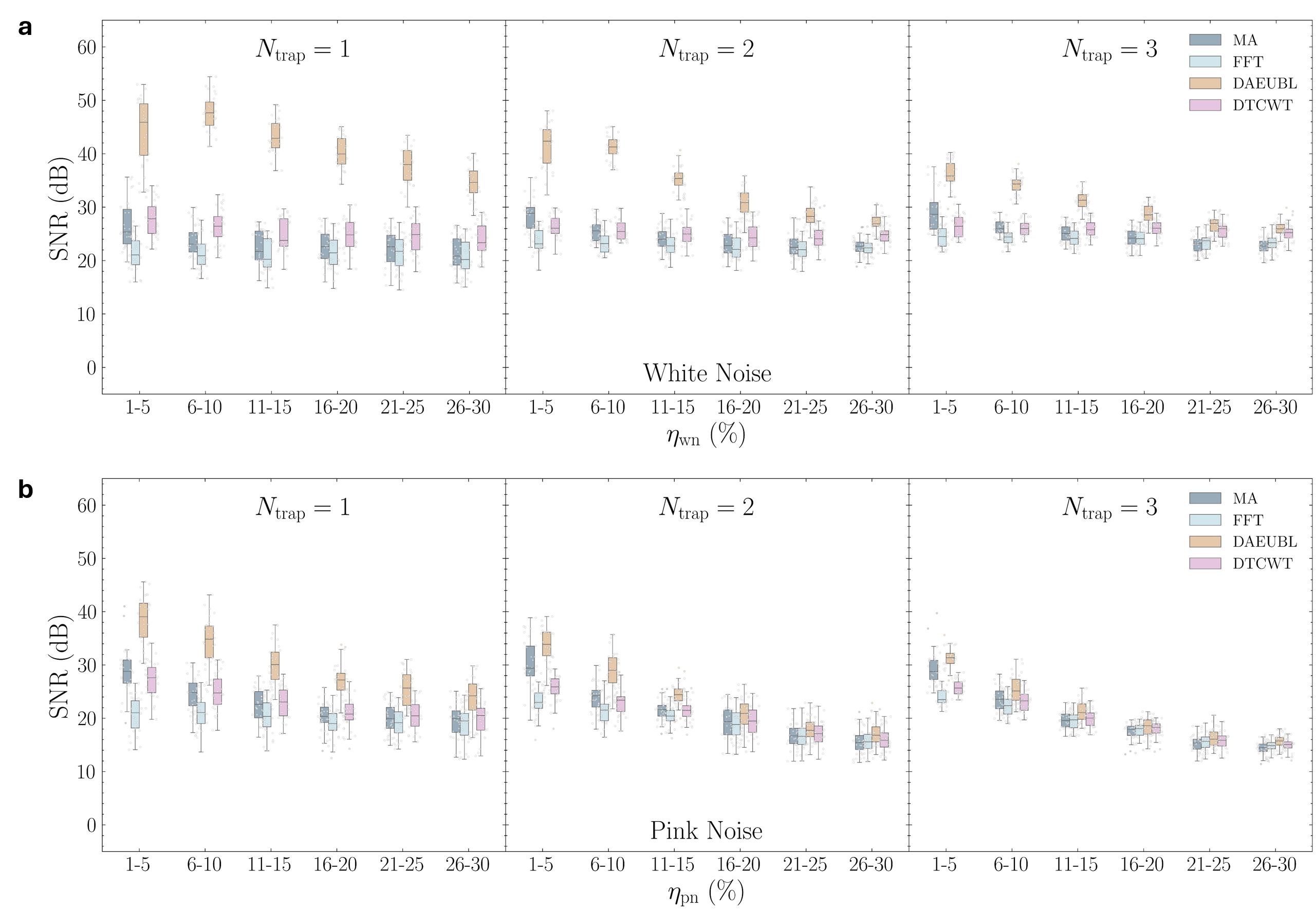}
    \caption{\textbf{Comparison of SNR (denoising quality) on benchmarked algorithms.} RTS samples spanning ground truth $N_{\text{trap}}=1,2,3$, and $\eta_{\text{wn}},\eta_{\text{pn}}=1\%\sim30\%$ for white \textbf{(a)} and pink noise \textbf{(b)}, respectively. RTS sample lengths fixed at $L = 100{,}000$ time steps.}
    \label{fig:fig2_snr}
\end{figure*}

\subsection*{Denoising Quality}

\subsubsection*{Signal-to-Noise Ratio (SNR)}

First, we evaluate the denoising performance using SNR as $\text{SNR (dB)} = 10 \cdot \log_{10} \left( {\sum_{t=1}^{L} x(t)^2}\big{/}{\sum_{t=1}^{L} (x(t) - \hat{x}(t))^2} \right)$, where $L$ is the length of the RTS signal in steps, $x(t)$ is the ground truth amplitude at time $t$, and $\hat{x}(t)$ is the denoised signal amplitude at the same time step. It is important to note that SNR may not fully reflect the capabilities of noise-removal methods such as DTCWT, particularly under pink noise conditions. This is because the \textbf{DTCWT + Bayesian} approach is designed to defer pink noise handling to the digitization stage via Bayesian inference. \autoref{fig:fig2_snr} plots the SNR in decibels (dB) over varying levels of white noise ($\eta_{\text{wn}}$, a) and pink noise ($\eta_{\text{pn}}$, b), across ground truth trap numbers $N_{\text{trap}} \in \{1, 2, 3\}$, respectively. As expected, SNR generally decreases with increasing noise levels ($\eta_{\text{wn}}, \eta_{\text{pn}}$) and higher trap complexity ($N_{\text{trap}}$). Moreover, lower SNR is observed under pink noise compared to white noise, as the more complex, non-white background noisy patterns significantly mask the RTS signals. Among four methods, DAEUBL achieves the highest SNR in all noise levels and trap numbers, corroborating its performance reported in the original work \cite{bowen.thesis}. The DTCWT method ranks second and yields SNR values above 20 dB in most cases, outperforming the other non-neural-network baselines (MA and FFT) across all tested conditions. Although DTCWT does not surpass the neural-network-based DAEUBL baseline in terms of SNR, this outcome is coherent with its design: the DTCWT primarily targets white noise and intentionally leaves pink noise components unfiltered for subsequent Bayesian post-processing. As a result, pink noise traces remain in the denoised signals, lowering the measured SNR. Nevertheless, DTCWT still achieves superior SNR compared to traditional MA and FFT baselines, while offering a more interpretable and efficient denoising strategy.

\begin{figure*}[!]
    \centering
    \includegraphics[width=0.95\linewidth]{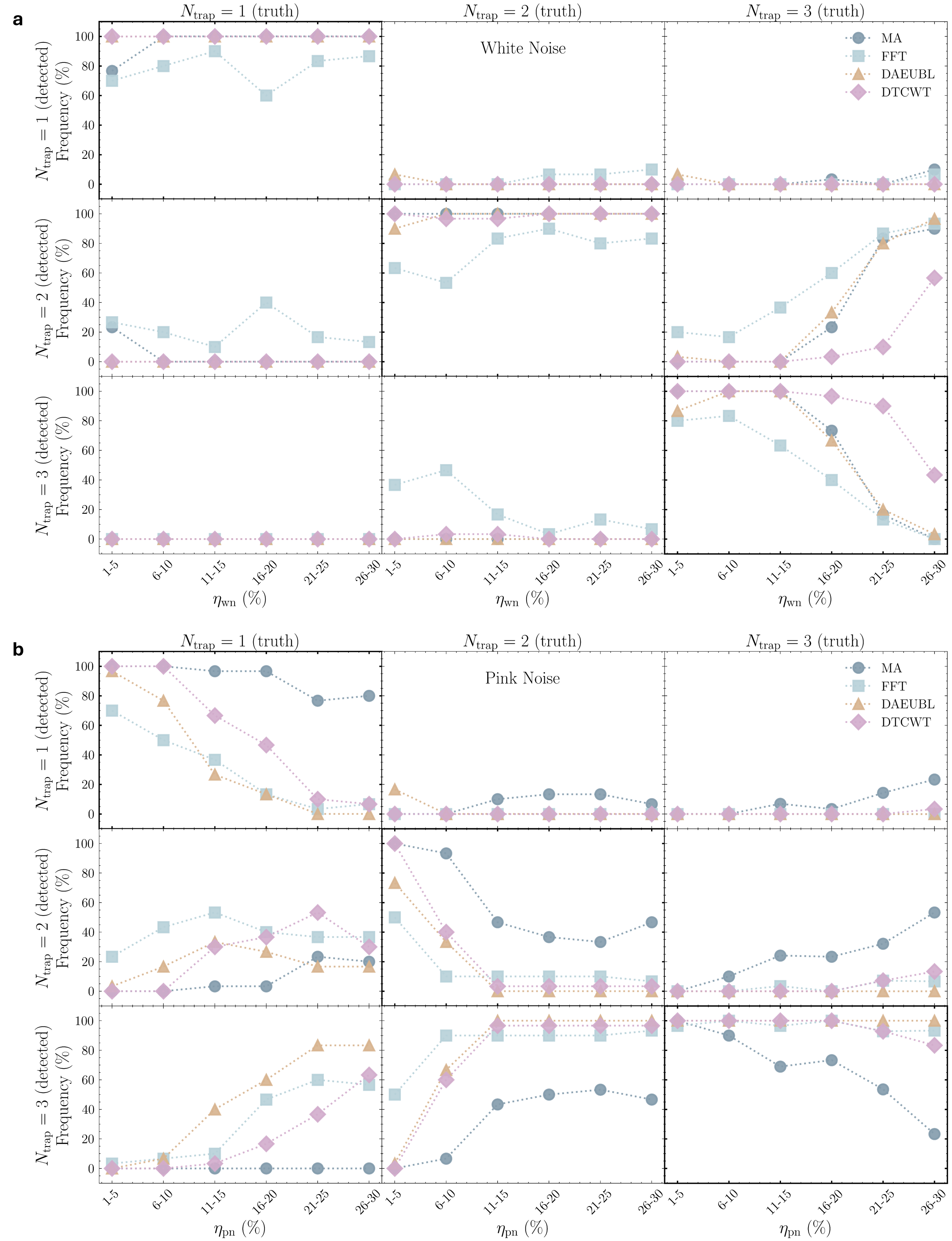}
    \caption{\textbf{Comparison of $N_{\text{trap}}$ error (denoising quality) on benchmarked algorithms.} RTS samples spanning ground truth $N_{\text{trap}}=1,2,3$, and $\eta_{\text{wn}},\eta_{\text{pn}}=1\%\sim30\%$ for white \textbf{(a)} and pink noise \textbf{(b)}, respectively. RTS sample lengths fixed at $L = 100{,}000$ steps. The darkened diagonal subplots indicate correct detections, namely, $N_{\text{trap, truth}} = N_{\text{trap, detected}}$. The values in each column of each method sum to 100\% of all RTS samples at the same ground truth $N_{\text{trap}}$ and $\eta_{\text{wn}},\eta_{\text{pn}}$ level. The bottom-left subplots indicate overestimation, while the top-right subplots reflect underestimation.}
    \label{fig:fig3_ntrap_error}
\end{figure*}

\subsubsection*{Trap Number Error}

At the validation step of our model, we evaluate the trap number detection accuracy by comparing $N_{\text{trap}}$ estimated from each method against the ground truth of the synthetic data. The detected trap number $N_{\text{trap}}$ is determined from the number of KDE levels in each denoised signal. Specifically, we compute 
$N_{\text{trap}} = \lceil \log_2(\text{No. KDE levels}) \rceil$, where the result is capped within the range $[1, 3]$ to align with the ground truth $N_{\text{trap, truth}}=1,2,3$ values used in this benchmark. \autoref{fig:fig3_ntrap_error} illustrates the trap number error results under white noise (\autoref{fig:fig3_ntrap_error}a) and pink noise (\autoref{fig:fig3_ntrap_error}b). Each subplot corresponds to a pairing of ground truth and detected $N_{\text{trap}}$: the three columns represent $N_{\text{trap, truth}}$ of 1, 2, and 3 traps, and the three rows represent $N_{\text{trap, detected}}$. Each subplot presents detection frequency on the y-axis and noise level on the x-axis. In the diagonal plots of the white noise in \autoref{fig:fig3_ntrap_error}a, our proposed method (adaptive DTCWT, shown in pink) outperforms all baselines 99\% of the time in correctly estimating $N_{\text{trap}}$ across $\eta_\text{wn}$ and $N_{\text{trap, truth}} \in \{1,2,3\}$. DAEUBL ranks second in most cases, and FFT always underperforms across all configurations. In practice, overestimation of $N_{\text{trap}}$ is generally less problematic than underestimation since extra levels from overestimation can be discarded based on prior physical knowledge of the system; however, underestimated $N_{\text{trap}}$ may miss meaningful trap events entirely. For the pink noise case in \autoref{fig:fig3_ntrap_error}b, trap number estimation becomes less stable, with stronger fluctuations in detection frequencies. Interestingly, MA performs best when $N_{\text{trap, truth}}=1,2$, but its effectiveness drops sharply at $N_{\text{trap, truth}}=3$, where it frequently underestimates. By contrast, DTCWT generally offers the most reliable performance, except at the highest $\eta_\text{pn} = 21\%\sim30\%$ noise levels for the $N_{\text{trap, truth}}=3$ case. Even in this challenging regime, our pipeline remains the top performer relative to other baselines. This stability reflects the ability of the DTCWT denoiser to suppress background noise while preserving transition features critical for KDE-based trap identification. DAEUBL, on the other hand, exhibits a clear tendency to overestimate the number of traps, which is particularly evident in the bottom row of subplots.


\subsubsection*{Trap Transition Amplitude Error}

The RTS trap transition amplitude ($\Delta_{\text{RTS}}$) is one of the three key parameters characterizing each RTS trap, alongside the high-state and low-state average dwell times ($\bar{\tau}_{\text{high}}$ and $\bar{\tau}_{\text{low}}$). $\Delta_{\text{RTS}}$ is estimated during the denoising and KDE step prior to digitization. Each KDE peak corresponds to a discrete signal level, which we represent using a binary index. In this representation, each bit indicates the state of a single trap: 1 = active (trap in high state), 0 = inactive (trap in low state). Interpreting the levels as least-significant-bit (LSB)-first binary codes allows us to map each observed level to a unique combination of trap states, from which $\Delta_{\text{RTS}}$ values for individual traps can be decoded. For example, in a system with two mutually independent traps, there can be up to four possible distinct levels: \texttt{00} corresponds to both traps being inactive, and \texttt{11} corresponds to both traps being active simultaneously. This scheme generalizes naturally: for $N_{\text{trap}}=3$, there are $2^3=8$ possible binary codes (\texttt{000}--\texttt{111}), each corresponding to a unique combination of trap states. The conversion is constrained by the number of traps detected ($N_{\text{trap}}$), ensuring that only valid binary indices are used in digitization. This decoding process effectively maps the continuous density peaks in KDE, shown on the left of \autoref{fig:fig1_rts_examples}c-h, into compact, interpretable binary representations. These signal levels allow us to compute the amplitude difference between each state transition and compare against the ground truth.

\begin{figure*}[!] 
    \centering
    \includegraphics[width=0.95\linewidth]{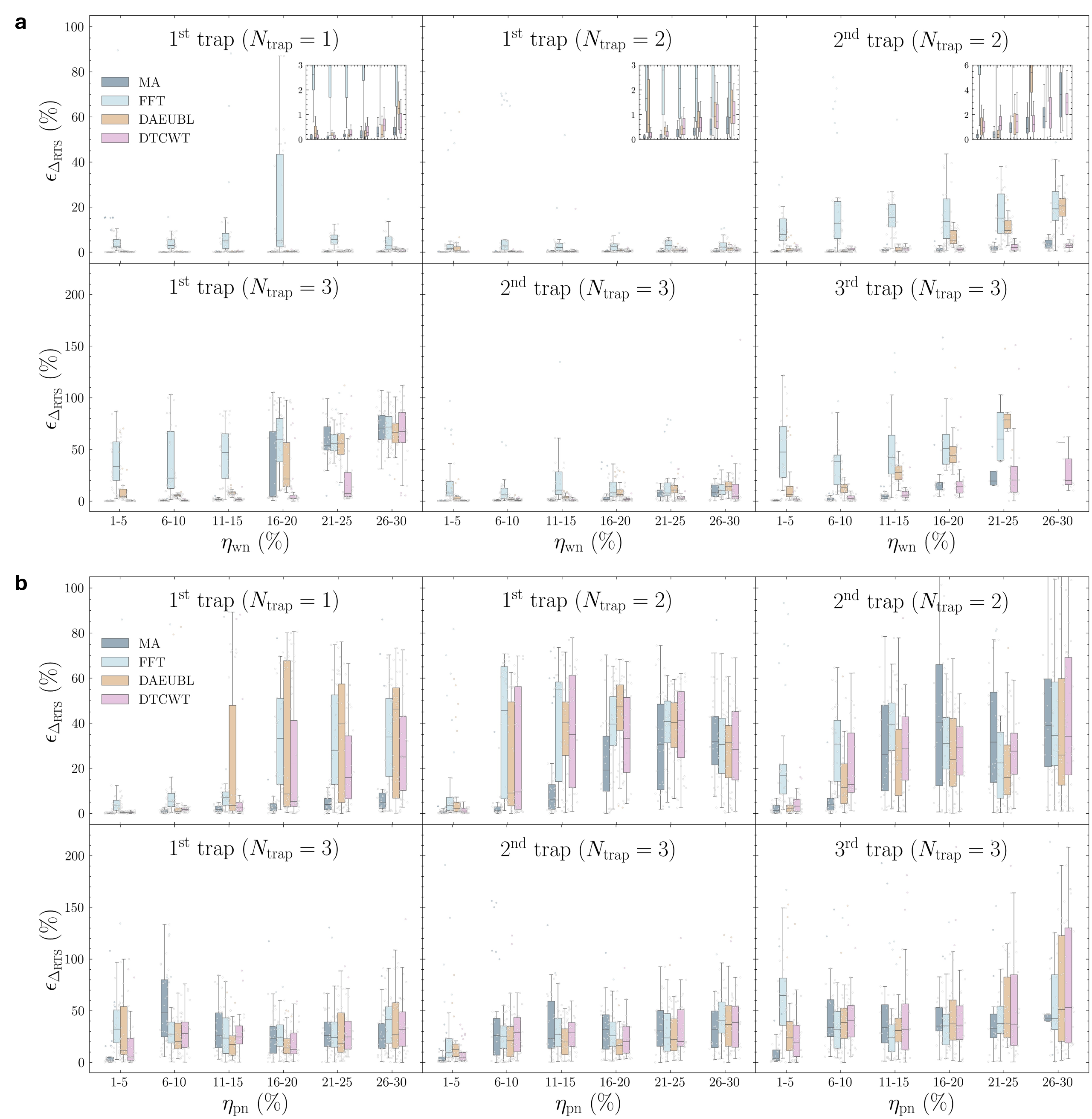}
    \caption{\textbf{Comparison of $\Delta_{\text{RTS}}$ error (denoising quality) on benchmarked algorithms.} RTS samples spanning ground truth $N_{\text{trap}}=1,2,3$, and $\eta_{\text{wn}}=1\%\sim30\%$ for white \textbf{(a)} and pink noise \textbf{(b)}, respectively. RTS sample lengths fixed at $L = 100{,}000$ steps. Error statistics of individual traps are separated into subplots. Each subplot contains results from up to 50 synthetic RTSs, provided the corresponding trap was correctly detected. The subplot layout is organized as follows: The top row, left subplot shows error distribution for the single trap in $N_{\text{trap}} = 1$ samples. The top row, middle and right subplots display errors for the two individual traps, labeled as the first and second traps, in $N_{\text{trap}} = 2$ synthetic RTSs. The bottom row shows errors for the first, second, and third traps in $N_{\text{trap}} = 3$ synthetic RTSs.}
    \label{fig:fig4_delta_rts_error}
\end{figure*}

We compute a percentage error $\Delta_{\text{RTS} }^{(i)}$ as $\epsilon_{\Delta_{\text{RTS}}}^{(i)}~(\%) = \left| (\Delta_{\text{RTS, detect}}^{(i)}-\Delta_{\text{RTS, truth}}^{(i)})~/~ \Delta_{\text{RTS, truth}}^{(i)}\right| \times 100 (\%)$ for each RTS detected $i^{\text{th}}$ trap, where $i \in \{1, 2, 3\}$. \autoref{fig:fig4_delta_rts_error} summarizes $\epsilon_{\Delta_{\text{RTS}}}^{(i)}$ as box plots of white noise (a) and pink noise (b) across different noise levels. Each subplot has the range of $\eta_\text{wn}$ levels on the $x$-axis and $\epsilon_{\Delta_{\text{RTS}}}^{(i)}$ in $y$-axis for the individual $i$-th trap. Across all trap configurations and noise levels, DTCWT consistently produces the lowest and most tightly concentrated $\Delta_{\text{RTS}}$ errors. The accuracy and stability of DTCWT in estimating trap amplitudes reinforce its robustness, especially under high noise conditions where other methods fail. The performance degradation of MA, FFT, and even DAEUBL at high noise levels aligns with previous observations from the $N_{\text{trap}}$ error analysis: when traps cannot be reliably identified, accurate $\Delta_{\text{RTS}}$ estimation becomes infeasible. For the pink noise case in \autoref{fig:fig4_delta_rts_error}b, $\epsilon_{\Delta_{\text{RTS}}}$ is generally higher than its white noise counterpart, with box plots exhibiting broader spreads that reflect greater variability in estimation. Unlike the white noise scenario, no single method emerges as a consistently dominant performer. Still, MA shows surprisingly strong accuracy in the $N_{\text{trap}} = 1$ setting, underscoring its effectiveness for simple single-trap signals even in the presence of structured pink noise.

\subsection*{Digitization Quality}

After completing the denoising and KDE stage, we proceed to evaluate the digitization quality in terms of two error metrics: trap state error as root mean square error (RMSE) of the binary digitized signal with respect to ground truth, and error in the dwell time statistics ($\bar{\tau}_{\text{high}}, \bar{\tau}_{\text{low}}$) for each trap. RMSE evaluates signal fidelity, while $\bar{\tau}$ statistics quantify transition behaviors. Among all metrics, we consider these RTS-specific characterization errors to be of higher priority than general signal metrics such as SNR, as they provide deeper insights into the RTS structure. More implementation and visualization details of these error metrics are presented alongside the respective plots in subsequent sections.

\begin{figure*}[!] 
    \centering
    \includegraphics[width=0.9\linewidth]{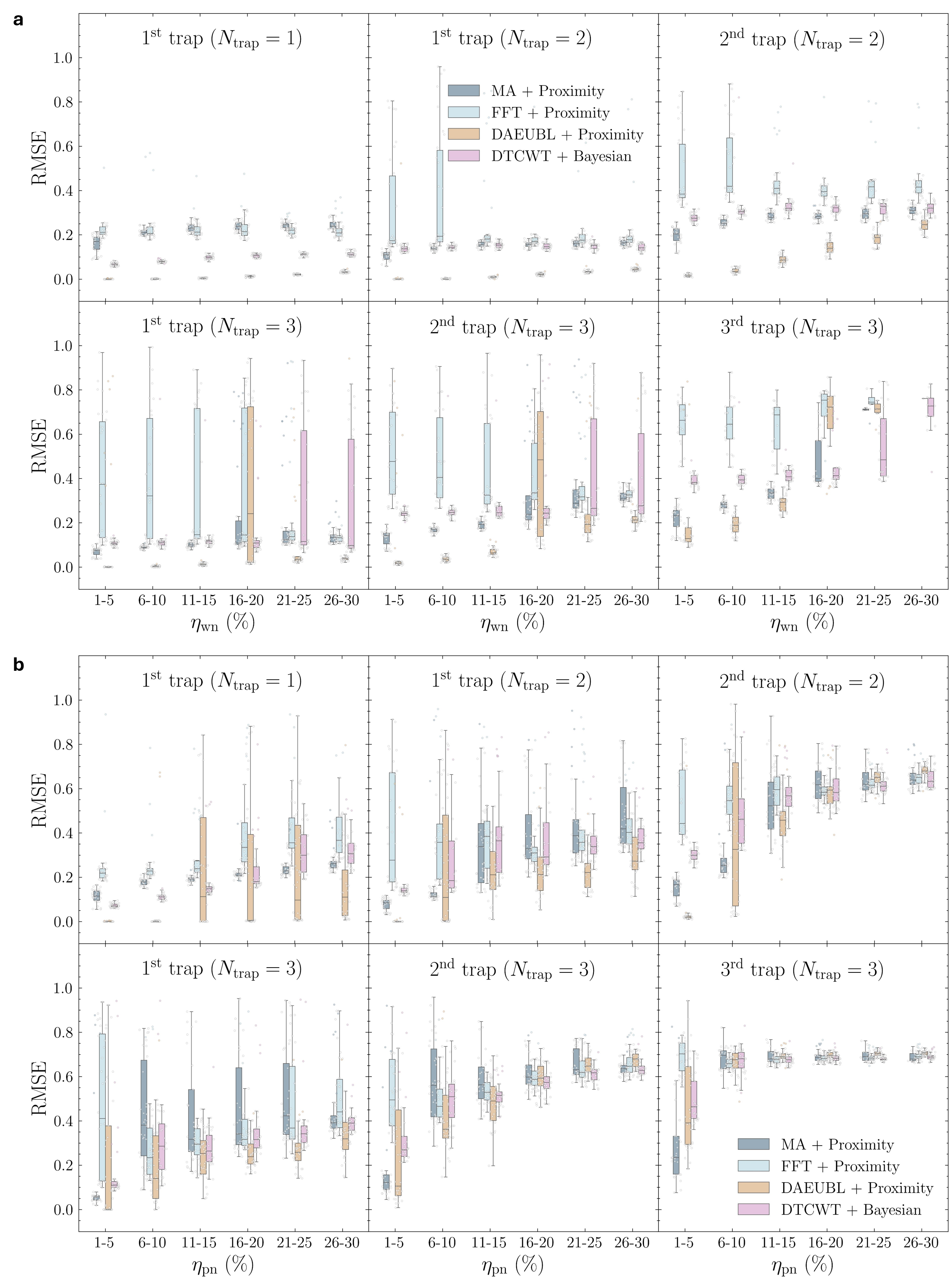}
    \caption{\textbf{Comparison of RMSE (digitization quality) on benchmarked algorithms.} RTS samples spanning ground truth $N_{\text{trap}}=1,2,3$, and $\eta_\text{wn},\eta_\text{pn} = 1\%\sim30\%$ for white \textbf{(a)} and pink noise \textbf{(b)}, respectively. Each RTS sample is fixed at a length of $L = 100{,}000$ time steps. Error statistics of individual traps are presented in separate subplots.}
    \label{fig:fig5_rmse}
\end{figure*}

\subsubsection*{Trap State Error}

 During the digitization step, one outcome is a digitized binary sequence for each trap, and the final digitized RTS signal is reconstructed by summing these sequences, each scaled by its respective $\Delta_{\text{RTS}}$ amplitudes. The first metric is the root mean squared error (RMSE) between the digitized binary sequences and the ground truth from the original synthetic data. Mathematically, the RMSE for the $i^{\text{th}}$ trap is then computed as $\text{RMSE}^{(i)} = \sqrt{ \frac{1}{L} \sum_{t=1}^{L} \left( b^{(i)}_{\text{digitized}}(t) - b^{(i)}_{\text{truth}}(t) \right)^2 }$, where $L$ is the RTS signal length in time steps, and $b^{(i)}(t) \in \{0, 1\}$ denotes the digitized state of the $i^{\text{th}}$ trap given $t$. Hence, this metric directly captures only state errors for each trap and is independent of amplitude deviations, unlike the RMSE computed for the full reconstructed signal. \autoref{fig:fig5_rmse} summarizes the RMSE results in the range of $[0, 1]$ along the $\eta_{\text{wn}},\eta_{\text{pn}}$ noise levels in (a) and (b), respectively. For the white noise results in \autoref{fig:fig5_rmse}a, while \textbf{DAEUBL + Proximity} shows the lowest RMSE in the $N_{\text{trap}} = 1$ and 2 cases,  it often fails to detect all three traps in $N_{\text{trap}} = 3$ samples, resulting in missing data for some subplots. It is important to note that these plots include only the traps that were successfully digitized. Therefore, the number of valid samples will differ across methods and noise levels. Our proposed method (DTCWT denoising + Bayesian digitization) demonstrates more consistent digitization success across all traps and noise levels, even when its RMSE is not always the lowest. In the presence of pink noise (\autoref{fig:fig5_rmse}b), our method generally yields the lowest and most concentrated RMSE values. Some exceptions appear in the $1^{\text{st}}$ trap of $N_{\text{trap}} = 1$ and 2 at very high noise levels ($\eta_{\text{pn}} = 26\%\sim30\%$), where \textbf{DAEUBL + Proximity} yields slightly lower median or second quartile in RMSE. Closer inspection of \textbf{DAEUBL + Proximity} box plots reveals significantly wider interquartile ranges (IQR) and pronounced outliers, indicating high variance and unreliability. Although the central tendency appears favorable, the variability suggests a lack of robustness compared to the more stable performance of our \textbf{DTCWT + Bayesian} method.

\begin{figure*}[!] 
    \centering
    \includegraphics[width=0.95\linewidth]{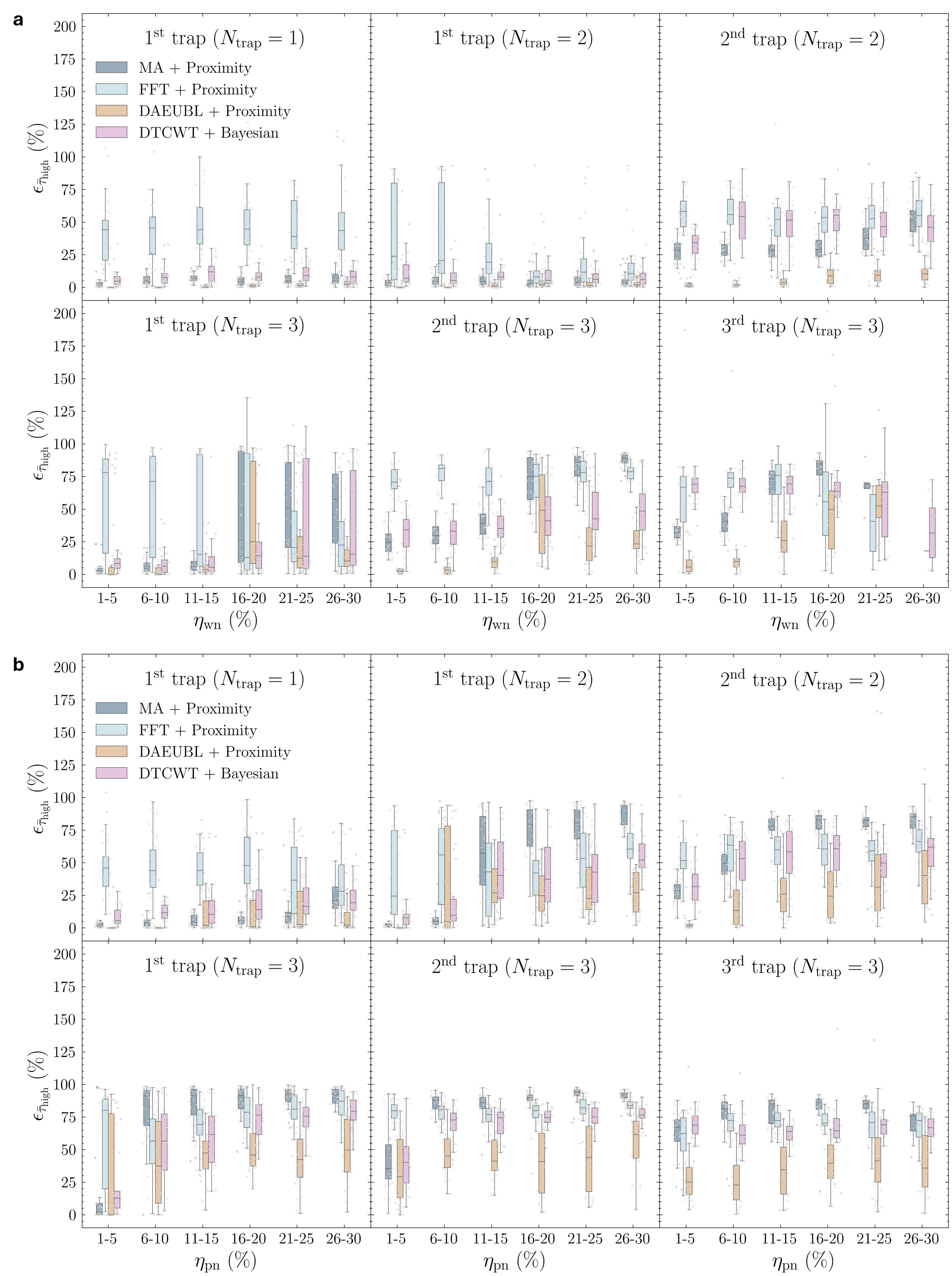}
    \caption{\textbf{Comparison of $\epsilon_{\bar{\tau}_\text{high}}$ (digitization quality) on benchmarked algorithms.} RTS samples spanning ground truth $N_{\text{trap}}=1,2,3$, and $\eta_{\text{wn}},\eta_{\text{pn}}=1\%\sim30\%$ for white \textbf{(a)} and pink noise \textbf{(b)}, respectively. RTS sample lengths fixed at $L = 100{,}000$ steps. Error statistics of individual traps are separated into subplots. Note that the box plots display individual points to present the cases where traps are successfully identified.}
    \label{fig:fig6_mean_tau_high_error}
\end{figure*}

\subsubsection*{Mean Dwell Times Error}

We next evaluate the accuracy of average dwell time estimation after digitization, quantified by the percentage error in the mean high-state and low-state dwell times, denoted as $\bar{\tau}_{\text{high}}$ and $\bar{\tau}_{\text{low}}$. For each individual trap in every RTS sample, $\bar{\tau}_{\text{high}}$ and $\bar{\tau}_{\text{low}}$ are extracted from the Poisson statistics fitting in the dwell-time histograms from the digitized signals. Note that the reference is the $\bar{\tau}_{\text{high}}$ and $\bar{\tau}_{\text{low}}$ from the same statistical analysis with. For the $i^{\text{th}}$ trap, the percentage error is calculated as $\epsilon_{\bar{\tau}^{(i)}_\text{high(low)}}(\%) = \left| {(\bar{\tau}^{(i)}_{\text{high(low), detected}} - \bar{\tau}^{(i)}_{\text{high(low), true}})/}{\bar{\tau}^{(i)}_{\text{high(low), true}}} \right| \times 100 \% $, where $i \in \{1, 2, 3\}$. \autoref{fig:fig6_mean_tau_high_error} collects the error distributions of $\bar{\tau}_{\text{high}}$ across noise levels for white noise (a) and pink noise (b). For $N_{\text{trap}} = 1$ and $N_{\text{trap}} = 2$ in \autoref{fig:fig6_mean_tau_high_error}a, the error grows as $\eta_\text{wn}$ increases for all models. In the $N_{\text{trap}} = 3$ case, \textbf{DAEUBL + Proximity} continues to perform well; however, at $\eta_{\text{wn}} = 26\%\sim30\%$, \textbf{DAEUBL + Proximity} and most other baselines frequently fail to detect the third trap. Only the proposed method (\textbf{DTCWT + Bayesian}) reliably captures the third trap under such challenging conditions. \textbf{DAEUBL + Proximity} can achieve lower error in favorable regimes; however, its reliability degrades more rapidly with increasing noise. By contrast, \textbf{DTCWT + Bayesian} maintains consistent detection and generally exhibits the smallest variation across error distributions, particularly under pink noise. A similar pattern is observed under white noise except for the first trap of $N_{\text{trap}} = 3$ at $\eta_{\text{wn}} = 21\%\sim30\%$, where variability increases. Under pink noise (\autoref{fig:fig6_mean_tau_high_error}b), the overall trends are consistent with the white noise scenario. \textbf{DAEUBL + Proximity} often yields the lowest median error, but this comes with a substantially wider IQR, reflecting larger variability. \textbf{MA + Proximity}, on the other hand, performs the weakest in most multi-trap configurations, despite showing stronger results in earlier metrics. The proposed method does not always achieve the lowest Q2, but it consistently produces tighter error distributions, reflecting greater stability. Thus, \textbf{DAEUBL + Proximity} excels in favorable conditions, whereas \textbf{DTCWT + Bayesian} is more robust and reliable across noise types and trap complexities. $\epsilon_{\bar{\tau}_\text{low}}$ also exhibits similar trends to $\epsilon_{\bar{\tau}_\text{high}}$; notably, only the proposed \textbf{DTCWT + Bayesian} method successfully captured the third trap in the $N{\text{trap}} = 3$ case at $\eta_{\text{pn}} = 26\%\sim30\%$.


\subsection*{Performance}

Beyond model quality, we assess the performance of denoising, digitization, and the overall processes in terms of two parameters: core execution time and peak memory usage. The datasets used for performance benchmarking consist of 240 RTS samples with varying lengths from $L = 100{,}000$ to $L = 20{,}000{,}000$ time steps. Although we also tested across different noise levels and numbers of traps, these factors did not show a significant impact on performance metrics. Thus, our performance benchmarks primarily focus on execution time and memory usage across different signal lengths. The following metrics are used for performance benchmarking: core execution time, capturing the runtime of each core algorithm (e.g., denoising, KDE, digitization), excluding file I/O operations. Then, peak memory usage occurs during the execution of each algorithm. All benchmarking experiments are performed under fixed hardware constraints to ensure fair and consistent comparisons. Specifically, all evaluations are conducted on a dedicated GPU node of the Compute Canada Beluga cluster, with 10 cores of Intel Gold 6148 Skylake CPU @ 2.4 GHz and 1 NVIDIA V100SXM2 GPU with 16 GB VRAM.


\begin{figure*}[!] 
    \centering
    \includegraphics[width=0.95\linewidth]{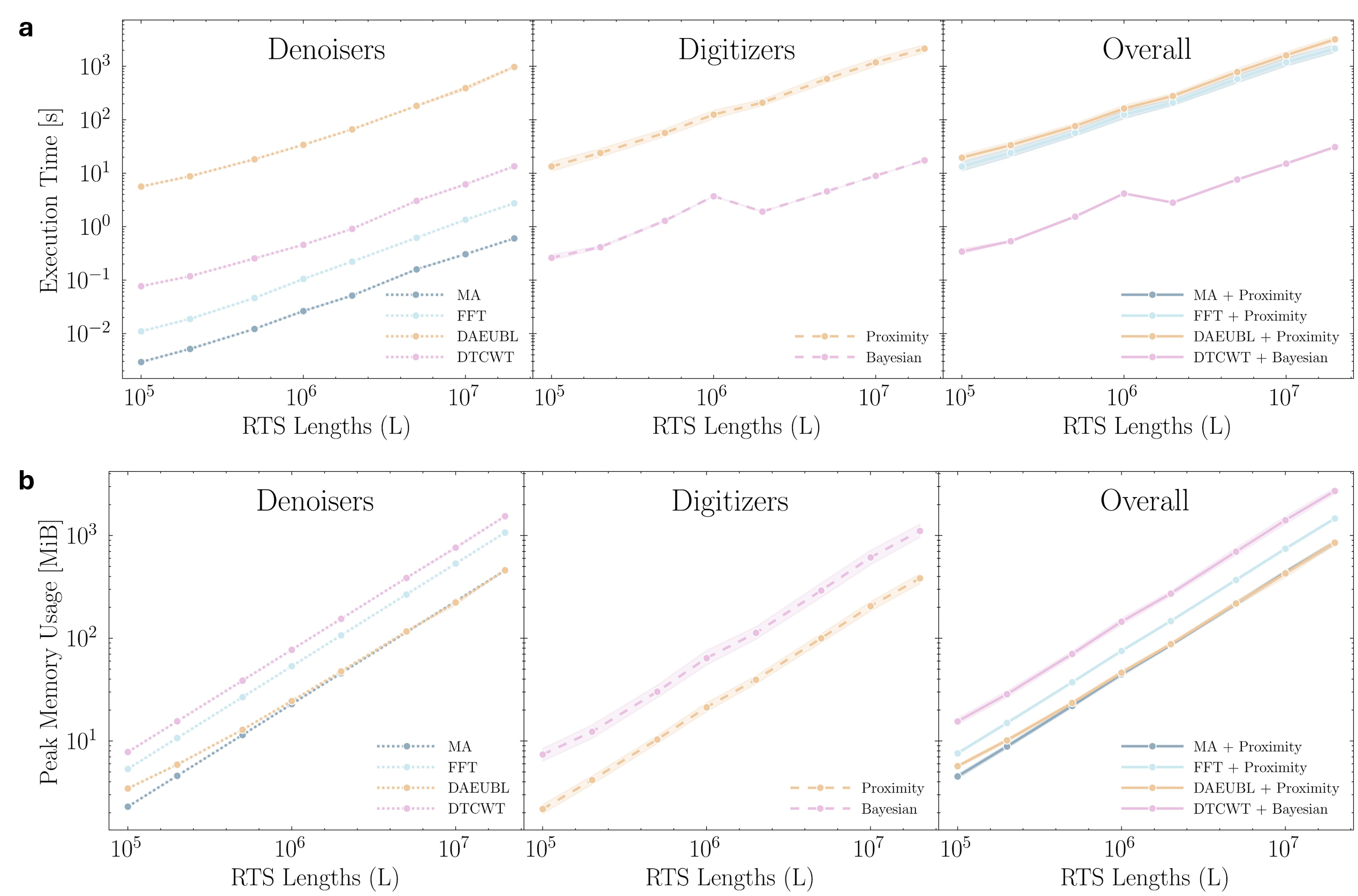}
    \caption{\textbf{Comparison of denoising and digitization performance on benchmarked algorithms. RTS samples spanning from $L = 100{,}000$ to $L = 20{,}000{,}000$ steps.} \textbf{(a)} Core execution time. \textbf{(b)} Peak memory usage. The shaded regions around the curves indicate the 95\% confidence intervals across RTS samples.}
    \label{fig:fig7_performance_benchmarks}
\end{figure*}

\subsubsection*{Core Execution Time}

The execution time for each core component of the pipeline is recorded individually for every RTS sample as a function of RTS length $L$. The plots in \autoref{fig:fig7_performance_benchmarks} therefore collect the results with a 95\% confidence interval band at each $L$. \autoref{fig:fig7_performance_benchmarks}a presents the execution time of denoising (left), digitization (middle), and the overall process (right) on a log-log scale. All four methods display a power-law trend as a function of $L$ (linear in log-log scale). The simpler methods (MA, FFT) take less time than the more complex methods (DAEUBL, DTCWT). Bayesian digitization outperforms the proximity method by nearly two orders of magnitude for all $L$ values in the middle subplot. Overall, it is evident that our proposed method, DTCWT denoising followed by Bayesian digitization is approximately 83 times faster than the other baseline approaches.

\subsubsection*{Peak Memory Usage}

\autoref{fig:fig7_performance_benchmarks}b plots the peak memory consumption in mebibytes (MiB) of each core component in the processing pipeline for every RTS sample in the performance evaluation dataset. Since DTCWT and Bayesian require higher peak memory usage than their counterparts, our proposed method, \textbf{DTCWT + Bayesian}, demands up to three times more peak memory compared to any baseline approach. Given that this configuration achieves an approximately 83-times improvement in execution speed, the associated increase in memory consumption represents a reasonable and justifiable trade-off.

\section*{Discussion}

Building on these benchmarking results, we now turn to a closer examination of the underlying mechanisms that drive the desirable performance of our proposed pipeline under challenging RTS conditions. First, the DTCWT denoiser exhibits the features of shift invariance, directional selectivity in the wavelet domain (i.e., distinguishing signal components by orientation in frequency rather than by time or amplitude), and strong noise suppression while preserving RTS step transitions. Unlike previous implementation requiring manual parameter tuning \cite{jackie.thesis}, our version includes an automatic parameter selection mechanism based on raw signal statistics, making it adaptive to diverse background noise regimes. For KDE, we employ a prominence-based peak filtering step that suppresses spurious peaks caused by noise, thereby improving the $N_{\text{trap}}$ estimation. This refinement is not the primary focus of our cross-method comparisons, since KDE is used consistently across all baselines, but it contributes to more reliable trap identification when it combines with the DTCWT denoiser. Second, the Bayesian digitizer replaces traditional heuristic-based digitization methods (e.g., thresholding or proximity rules) with a probabilistic latent-level model over KDE peaks. For each time step, we compute the full posterior distribution under a Gaussian emission model and then report the digitized label via maximum a posteriori (MAP) decoding. Along with temporal prior smoothing, this yields a principled measure of uncertainty, reduces susceptibility to noise, and naturally extends to multi-trap scenarios where interactions between levels complicate deterministic digitization. This allows for more reliable digitization and scales well in multi-trap settings. The resulting binary or multi-level sequences enable the best possible quantification of physical RTS parameters, $\bar{\tau}_{\text{high}}$, $\bar{\tau}_{\text{low}}$ and $\Delta_{\text{RTS}}$.

With increasing demands for real-time RTS analysis in applications such as quantum random number generation (QRNG) and semiconductor quality control during fabrication, efficient signal processing is becoming a necessity. Our proposed method demonstrates practical real-time performance: 20 million steps of RTS data can be processed in just 30 seconds, which corresponds to 1 second of signal processed in 1 second if sampled at 667 kHz (1$\mu$s time bin). In contrast, the DAEUBL-based pipeline requires about 3000 seconds to process the same data, supporting only up to 6.67 kHz resolution for real-time usage, even if we exclude the training time of the DAEUBL model. Regarding memory, our evaluation focuses on peak memory usage, which measures the maximum memory consumption during the entire execution. This metric is particularly relevant for future deployment scenarios, such as implementing RTS analysis pipelines on embedded or standalone hardware where memory is often a critical constraint. Ensuring that peak memory remains within feasible limits is essential for such implementations. It is important to contextualize the observed memory footprint. DAEUBL was highly optimized using Keras, and classical denoisers such as MA and FFT were implemented with efficient libraries like NumPy and SciPy. In contrast, our DTCWT implementation has not undergone memory optimization yet. Nevertheless, the observed increase in peak memory, up to approximately three times higher than other methods, is a reasonable trade-off considering the 83 times improvement in execution speed. This suggests that future optimizations of DTCWT memory usage could further enhance its suitability for low-resource environments without compromising speed.

\begin{figure*}[!] 
    \centering
    \includegraphics[width=0.95\linewidth]{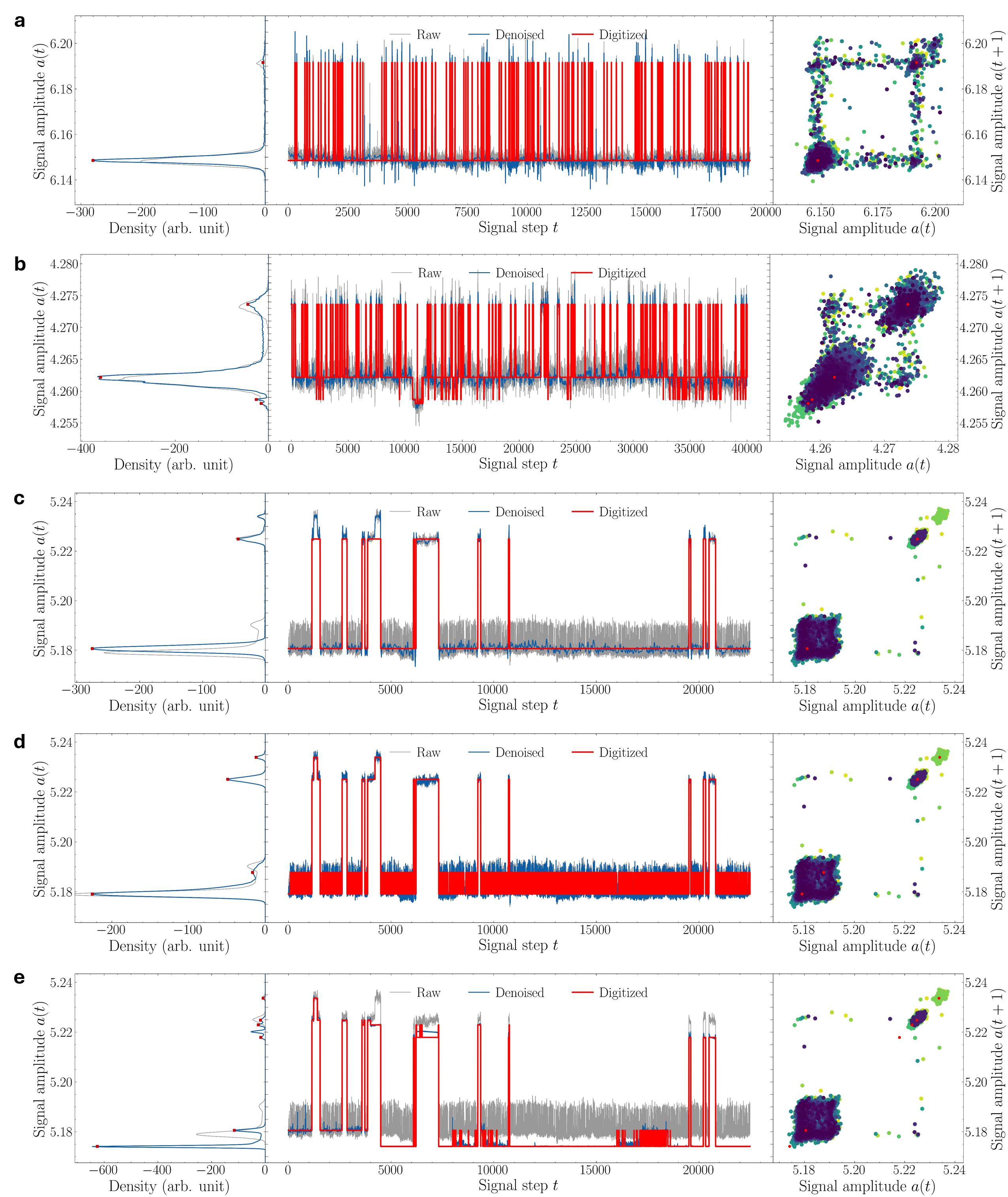}
    \caption{\textbf{Examples of processed real RTS from carbon nanotube film} with noisy RTS (grey), denoised RTS (blue), and digitized RTS (red) on middle subplot, cropped for better visualization; kernel density estimation (KDE) plot on left subplot; time-lag plot on right subplot for the entire RTS duration. \textbf{(a-c)} RTS from three distinct samples (Samples 1–3) processed using fully autonomous DTCWT + Bayesian. \textbf{(d)} RTS from Sample 3 reprocessed using manually tuned DTCWT + Bayesian following expert recommendation on potential alternative trap behavior. \textbf{(e)} RTS from Sample 3 reprocessed using DAEUBL + Proximity.}
    \label{fig:fig8_real_rts}
\end{figure*}

Beyond quantitative evaluation, the proposed approach offers several key practical advantages. Most notably, it does not require training of machine learning models, enabling it to generalize across a wide variety of RTS conditions without overfitting to specific data distributions. In contrast, neural-network-based methods like DAEUBL, although being powerful under ideal conditions, are often less reliable on unseen real RTS data and demand substantial expertise to train, validate, and deploy \cite{jackie.thesis}. Our method is inherently modular, allowing users to plug in different stages if needed, while remaining robust with minimal intervention. We also evaluated our method on experimental RTS signals collected from a carbon nanotube (CNT) device, visualized in \autoref{fig:fig8_real_rts}. The device under test consists of a 500 nm-wide, 65 nm-thick multi-walled CNT film acting as a conduction channel, with two terminals for voltage bias. The current through the channel is recorded at a temperature of 9 K. We emphasize that these real-data examples are presented to demonstrate practical applicability rather than to provide a complete physical characterization of CNT trap dynamics, which typically requires device-specific studies such as gate-voltage or temperature dependence and is beyond the scope of this work. In \autoref{fig:fig8_real_rts}a, the time-lag plot of this sample shows mostly round-shaped intensities, indicative of white noise. The RTS signal is clean and clearly exhibits discrete transitions, making it well-characterized as a single-trap RTS. The denoiser performs smoothly and confidently classifies the step levels. The sample in \autoref{fig:fig8_real_rts}b exhibits stretched intensities in the time-lag plot, consistent with pink noise. Although it is noisier, it remains interpretable. The KDE reveals two lower peak levels near 4.258. Depending on tuning, one could retain one of these levels to model the signal as a 2-trap RTS with three levels, or discard both and interpret it as a simplified single-trap RTS. Such ambiguity reflects common experimental scenarios where expected trap behavior may depend on prior physical insight; the proposed pipeline provides a reasonable autonomous interpretation while allowing expert-guided refinement when desired. The last sample in \autoref{fig:fig8_real_rts}c also displays mostly white noise characteristics in the time-lag plot. Although the KDE suggests a potential high-level peak, the denoised signal offers limited support for this level, particularly from $t=0$ to $t=5000$. The time-lag density confirms that this upper peak occurs much less frequently, justifying its exclusion under fully autonomous operation. However, prior literature \cite{rts.SR} has suggested that the fast switching behavior below approximately 5.20 may correspond to a separate trap; motivated by this, we include an additional tuned result in \autoref{fig:fig8_real_rts}d, where reduced denoising strength preserves these transitions and yields a multi-trap interpretation. This comparison illustrates a key strength of the proposed framework: it enables reliable autonomous analysis across diverse noise regimes, while remaining flexible enough to incorporate expert expectations about trap behavior when such knowledge is available. In contrast, DAEUBL shows reduced robustness on this unseen real RTS example \autoref{fig:fig8_real_rts}e, as its training and noise synthesis do not fully span the experimental conditions encountered here.

Our solution is training-free, fast, generalizable, and easy to use for non-ML experts. While no single approach can dominate across all RTS types due to their inherent variability, our framework comes close by balancing accuracy, robustness, and usability. This leads to our future direction: building a fully adaptive RTS analysis pipeline that can automatically select optimal methods at each stage of the workflow. Rather than benchmarking all algorithm combinations on every sample, such a system would dynamically recommend the best configuration, balancing accuracy and computational cost—based on real-time signal characteristics. This will support a more scalable and intelligent RTS characterization protocol applicable to both synthetic and real-world datasets, fulfilling the long-standing need for a standardized, high-throughput RTS analysis toolkit. Development of new denoisers and digitizers remains central to this goal. An alternative direction involves direct digitization, segmenting the noisy signal into tens or even hundreds of candidate traps, where high-frequency, short-lived transitions are interpreted as noise and discarded. This approach could retain only the dominant, stable trap states, potentially preserving accurate dwell statistics without requiring explicit denoising. In addition, neural networks offer promising avenues for RTS digitization beyond classical rule-based methods. While most existing neural approaches focus on denoising with interpretable outputs, lightweight architectures could be trained to directly perform digitization, learning to distinguish genuine trap transitions from noise. Future studies will explore such strategies to further enhance flexibility and performance in RTS analysis.


\section*{Methods}

\subsection*{Generator Protocol}

A synthetic signal generation pipeline is constructed for mimicking the physical switching behavior of RTS traps, which enables the quantitative evaluations of denoising methodologies, such as the number of traps $N_{\text{trap}}$ and various noise environments, with noise level $\eta_{\text{wn}}$ or $\eta_{\text{pn}}$ corresponding to white noise or pink noise type. Starting from the ground truth RTS signal, we progressively corrupt it with controlled noise, enabling precise benchmarking of denoising and digitization algorithms. Each RTS trap is modeled as a two-state (high/low) Markov process with probabilistic switching governed by characteristic lifetimes $\bar{\tau}_{\text{low}}$ and $\bar{\tau}_{\text{high}}$. These correspond to the average durations that the system stays in the low and high states, respectively, and are linked to physical trapping/de-trapping mechanisms in nanoscale devices. We synthesize each trap component as a binary sequence characterized by three parameters: $\bar{\tau}_{\text{low}}$, $\bar{\tau}_{\text{high}}$ with each switching durations sampled from a geometric distribution, and a corresponding trap amplitude defined by $\Delta_{\text{RTS}}$. Multiple such components are synthesized and summed to construct a ground truth multi-trap signal, which is equivalent to the physical superposition in multi-trap RTS.

Traps are assumed to be independent in order to provide a controlled baseline for benchmarking that isolates algorithmic performance without introducing device-specific trap interactions. While correlated traps may arise in certain experimental systems, the independent-trap assumption does not trivialize the analysis task. To emulate realistic experimental conditions, controlled noise is added to the ground truth RTS signal, which can be either single-trap or multi-trap. $A(t)$ denotes the ground truth RTS amplitude at time or signal step $t$, to which noise is added with intensity scaled by a factor $\eta_{\text{wn}}$. Let $\Delta A = \max(A) - \min(A)$ be the RTS maximum amplitude range. Upon synthesizing the white noise to the ground truth signal, we define the white noise $W(t) \sim \mathcal{N}(0, \sigma_W^2)$ where $\mathcal{N}(0, \sigma_W^2)$ denotes a Gaussian distribution with zero mean and variance $\sigma_W^2$. The standard deviation is $std(W(t))=\sigma_W = \Delta A \cdot \eta_{\text{wn}}$. The synthesized RTS with white noise is then expressed as $a(t) = A(t) + W(t)$. To synthesize RTS samples with pink noise as a form of low-frequency noise, we replace the previously modeled Gaussian white noise with pink noise, characterized by a power spectral density (PSD) that follows a $1/f^\alpha$ trend, where $f$ denotes frequency and $\alpha$ is the power-law exponent. We assume the standard pink noise with $\alpha = 1$. The pink noise $P(t)$, when the given noise level is $\eta_{\text{pn}}$, is derived by applying a power-law filter to white noise $W(t)$ where $\eta_{\text{wn}}=\eta_{\text{pn}}$. The filtered output $\text{PL}(t)$ has a $1/f$ PSD characteristic. Therefore, the pink noise term is defined as $P(t) = \sigma_W \cdot \text{PL}(t)$, where $\text{PL}(t)$ is a pink noise sample with \(1/f\) spectral properties by first generating white noise. The resulting pink noise RTS thus becomes $a(t) = A(t) + P(t)$. The process of adding controlled noise can be visualized in \autoref{fig:fig1_rts_examples} from ground truth RTS (black) to noisy RTS (grey). Although the noise is additive, the inclusion of strong low-frequency (pink) noise introduces substantial temporal correlations and background fluctuations that can fully mask weaker RTS transitions, reflecting a major challenge commonly encountered in experimental RTS measurements.

All stochastic procedures, including RTS switching and noise generation, are seeded using a signal-specific random seed. Given the RTS parameters and seed, the exact signal can be regenerated. This deterministic setup ensures traceability and supports debugging and reproducible experimentation. In summary, the RTS generator synthesizes white noise or pink noise RTS based on sample length $L$, number of traps $N_{\text{trap}}$, corresponding noise level $\eta_{\text{wn}}$ or $\eta_{\text{pn}}$, as well as a random seed for reproducibility. This modular pipeline provides a flexible framework for generating synthetic RTS signals suitable for evaluating denoising algorithms in a variety of noise regimes.

\subsection*{Denoising and KDE Algorithm}

The denoising module takes a raw amplitude time-series signal as an input, which typically exhibits substantial background noise that can mask the discrete switching events characteristic of RTS behavior. The resulting denoised signal closely approximates the idealized RTS form and provides a more interpretable basis for further processing stages. An example of this transformation from noisy RTS (grey) to denoised RTS (blue) is illustrated in \autoref{fig:fig1_rts_examples}c-h. The moving average (MA) filter is simple and fast \cite{RTS_studies1, RTS_studies2}, with an adaptive window size $W$ using the relation $W = \left\lfloor 60 \cdot \tan^{-1} (\sigma_{\text{wn}}/4)/\pi\right\rfloor$, where $\sigma_{\text{wn}}$ represents the standard deviation of the background white noise, balances noise suppression against transition preservation, though it remains sensitive to noise characteristics. FFT-based denoising operates in the frequency domain by retaining only the top 1\% of spectral coefficients, effectively reducing broadband noise but risking artifacts when signal and noise spectra overlap. Finally, the DAEUBL neural model provides strong performance under pink noise and multi-trap scenarios, but at significant computational and memory cost. 

Beyond adapting the DTCWT for RTS denoising, a central innovation of our approach lies in the development of two autonomous parameter selection rules that automatically determine (i) the optimal number of wavelet decomposition levels and (ii) the threshold value for wavelet coefficient pruning. These parameters are essential for achieving a balance between preserving abrupt switching behavior and effectively suppressing both white and colored background noise. Unlike prior studies \cite{jackie.thesis} that relied on manual tuning or trial-and-error, our method eliminates this manual burden by using data-driven heuristics derived from empirical testing across, including synthetic RTS and real RTS from quantum dot measurements, single-photon avalanche diode signals. If the user does not provide these parameters explicitly, the denoiser invokes the internal selection rules, as outlined in \autoref{alg:dtcwt_denoising}. After thresholding, the signal is reconstructed by combining retained features across all decomposition levels, resulting in a clean, denoised signal that maintains step fidelity under diverse noise conditions. These autonomous parameter selection rules serve a similar purpose to previous automation efforts in MA filtering, where the window size was adaptively adjusted based on noise-level estimates \cite{bowen.thesis,rts.SR}.

\begin{algorithm}[H]
\caption{Adaptive DTCWT Denoising}\label{alg:dtcwt_denoising}
\begin{algorithmic}[1]
\Require Normalized RTS, optional: decomposition levels, threshold
\Ensure Denoised signal
\If{decomposition levels not provided}
    \State Estimate levels using signal length and log-scaling rule
\EndIf
\If{threshold not provided}
    \State Compute spectral entropy of the signal
    \State Estimate threshold based on entropy value
\EndIf
\State Perform forward DTCWT on input signal
\For{each highpass level}
    \State Zero out values below threshold
\EndFor
\State Replace thresholded highpasses in the wavelet result
\State Perform inverse DTCWT to reconstruct the denoised signal
\State \Return Denoised RTS
\end{algorithmic}
\end{algorithm}

Since the DTCWT denoising requires both a decomposition level and a threshold value, we conduct an exhaustive grid search over reasonable parameter ranges to identify optimal combinations. This procedure is performed on a small set of representative real RTS measurements for which reference estimates were available, and is not based on the synthetic benchmark data used for evaluation. For each RTS signal, the data are duplicated across the full parameter grid and processed through DTCWT denoising followed by digitization. We then compute the deviation between the extracted and reference mean dwell times ($\bar{\tau}_{\text{low}}$, $\bar{\tau}_{\text{high}}$), and identify parameter regions that jointly minimize errors in both states. From these experiments, we observe strong statistical correlations between the optimal DTCWT parameters and simple measurable properties of the raw RTS signal. Importantly, these relationships depend only on general signal statistics and not on semiconductor-specific operating conditions or device physics. Specifically, we find that the optimal decomposition level $K$ correlates with the signal length $L$, while the optimal wavelet coefficient threshold $T$ is best predicted by the spectral entropy $H_S$ of the signal. This enables us to derive the following predictive rules using linear regression fits over the grid search results:

\begin{itemize}
    \item $K = 0.59 \cdot \log_2(L) - 4.05 \quad (R^2 = 0.95) ,$
    \item $T = -21.40 \cdot H_S + 137.15 \quad (R^2 = 0.99) .$
\end{itemize}

Here, $H_S$ is defined as $H_S = -\sum_{k=1}^{N} P_k \log P_k$, where $N$ is the total number of frequency bins obtained from Welch's method, $X_k$ is the power spectral density coefficient at the $k$-th frequency bin, and $P_k = |X_k|^2 / \sum_{i=1}^{N} |X_i|^2$ corresponding normalized power spectral density of the signal. Spectral entropy was found to provide the most consistent correlation with the optimal threshold among the signal statistics examined. The resulting rules enable automated selection of DTCWT denoising parameters directly from the raw signal, without requiring prior knowledge of the underlying semiconductor system. The automatic selection does not fail in the sense of halting the algorithm or producing undefined outputs; rather, under extreme noise conditions, limitations manifest as increased benchmark errors, reflecting fundamental RTS reconstruction limits shared by all denoising methods. For real experimental RTS signals, where no ground truth exists, these rules provide a principled and reproducible starting point that can be further refined by expert users if desired. For clarity in this study, we refer to our fully automated DTCWT-based denoiser with adaptive selection of $K$ and $T$ simply as \textbf{DTCWT}, distinguishing it from the original transform and from basic denoising implementations that require explicit parameter specification.   

Although we benchmark against the DAEUBL approach, our noise injection protocol differs significantly from that in the DAEUBL proposal \cite{bowen.thesis}, which applies the noise ratio relative to the minimum $\Delta_{\text{RTS}}$ across all traps, whereas our method uses the maximum $\Delta$ from the amplitude span of the full multi-trap RTS signal as the reference for scaling noise. This results in a fundamental difference in how the same nominal $\eta_{\text{wn}}$ or $\eta_{\text{pn}}$ translates into actual noise magnitude in the signal. For instance, consider a signal with three traps where the individual trap steps are: 0.2, 0.3, and 0.4 units. DAEUBL would base its noise scaling on the smallest step (0.1), while our approach uses the full range (0.9 if all traps are on). Therefore, a $\eta_{\text{wn}}=20\%$ noise in our system ($\sigma = 0.9 \times 20\% = 0.18$) would correspond to a 90\% noise level under DAEUBL proposal ($0.18 / 0.2 = 90\%$) \cite{bowen.thesis}. This discrepancy should be considered when comparing denoising performance metrics across studies, as the apparent signal quality at a given noise ratio is not directly equivalent between reported studies.

After denoising, the next step in the pipeline is to identify the stable amplitude levels present in the signal. For this purpose, we apply KDE, a non-parametric technique used to estimate the probability density function of the signal amplitudes. KDE operates by placing a smooth kernel, typically a Gaussian function, over each data point in the denoised signal, and summing the resulting distributions to form a continuous density function. This approach generates a smooth approximation of the amplitude distribution, which is particularly well-suited for noisy or fluctuation-heavy signals like RTS. Because KDE peak detection is sensitive to bandwidth selection, we use a fixed rule-of-thumb bandwidth to ensure reproducibility across all benchmark signals. Initially, KDE is applied to one-dimensional RTS amplitude samples ($d=1$), with the signal length $L$ corresponding to the number of samples. While Scott’s rule would prescribe a bandwidth scaling factor of $L^{-1/(d+4)} = L^{-1/5}$, we deliberately replace this data-dependent factor with a fixed empirical bandwidth (bw $=0.020$) applied uniformly after denoising and amplitude normalization. We found that using a fixed bandwidth yields more stable and interpretable KDE peak structures than bandwidths that shrink with signal length $L$. The value 0.020 was selected empirically as a conservative smoothing scale that balances suppression of residual fluctuations with preservation of distinct RTS levels. Within a reasonable range around this value, the number, ordering, and prominence of dominant KDE peaks remained stable in our tests. By avoiding signal-length–dependent bandwidth shrinkage inherent to rule-of-thumb methods such as Scott’s rule, this choice ensures consistent peak resolution across signals of varying length $L$ and noise level $\eta_{\text{wn}}$ or $\eta_{\text{pn}}$ without per-signal tuning. Peak identification in the KDE result is performed using prominence-aware peak-finding algorithms. These methods assess not only the height of peaks but also their relative separation from neighboring fluctuations, providing robustness in noisy scenarios. When analyzing known 1-trap signals, it is possible to apply a forced mode to extract exactly two dominant peaks, ensuring that the digitization process focuses on the most physically meaningful states. However, this is not used throughout the entirety of this study to ensure no intervention in executing the protocols, meaning that it is possible to recognize a 1-trap RTS as a multi-trap when too much residual noise is present after denoising. Compared to traditional histograms, which can produce misleading visualizations due to arbitrary bin widths or boundaries, KDE offers a smoother and more faithful representation of the underlying distribution. This is especially useful when identifying discrete RTS levels, as noise and sampling artifacts can otherwise obscure the presence of multiple states. In the context of RTS analysis, KDE enables clearer visualization and quantification of distinct amplitude levels—corresponding to the discrete states of traps. These appear as peaks in the KDE output. For example, a 1-trap RTS typically produces two prominent peaks, corresponding to the high and low states. These peak positions are then used in the digitization step to assign each sample in the signal to its most probable discrete state. As illustrated in \autoref{fig:fig1_rts_examples}c–h, the left subplots show the KDE results of the denoised RTS signals, where the density curves reveal the dominant peaks used to determine trap amplitude levels. The middle subplots then illustrate the progression from denoised RTS (blue) to digitized RTS (red). Unlike a histogram, which may display spurious gaps or bins that obscure these levels, the KDE plot offers a continuous and interpretable view that emphasizes the underlying state levels. This clarity is critical for downstream digitization and for accurate extraction of switching behavior. We emphasize that KDE peak detection is intended to provide an automated and reproducible starting point for RTS analysis, while KDE may not perfectly identify all RTS levels in particularly challenging cases, such as those with extreme noise. These limitations are consistent with fundamental signal reconstruction constraints. In practice, such difficult experimental RTS traces often require expert inspection or manual refinement, which remains standard in experimental workflows. Accordingly, achieving zero error is neither expected nor required for the benchmarking goals of this study, and the impact of KDE uncertainties is explicitly captured by the proposed benchmarking metrics.

\subsection*{Digitization Algorithm}

In the digitization step, each point in the denoised signal is assigned to discrete amplitude levels detected from KDE peaks, corresponding to distinct multi-trap states (i.e., combinations of individual binary traps). Each continuous-valued amplitude sample is mapped to its nearest KDE peak, effectively converting the signal into a stepwise sequence of discrete levels. These levels represent the most likely trap state configurations at each point in time. The outcomes of this process are illustrated as noisy RTS (gray), denoised RTS (blue), and digitized RTS (red) in \autoref{fig:fig1_rts_examples}c-h. The quality of both the denoising and digitization stages can be qualitatively assessed by comparing the digitized trace against the ground truth RTS (black). In the simplest case of a single trap, the digitized signal reduces to a binary time series alternating between two distinct levels, reflecting the “on” and “off” occupancy states. However, as the number of traps increases, the RTS becomes multi-level: a two-trap system, for example, yields four unique amplitude states corresponding to combinations of the two binary trap configurations (00, 01, 10, 11). In such cases, accurate digitization becomes increasingly challenging due to overlapping levels and closely spaced transitions. Several digitization strategies are employed to address this complexity, each with trade-offs in robustness and computational demand. One of the simplest approaches is proximity-based level assignment, extensively used in prior studies \cite{bowen.thesis,rts.SR}, where each amplitude point is matched to the closest KDE peak. This method is fast and effective for low-noise, single-trap signals, where transitions are well-separated. However, it becomes error-prone in high-noise or multi-trap contexts, where peak overlap and noise artifacts can lead to misclassification, spurious transitions, and distorted dwell statistics. An HMM-based digitizer was also implemented as a variant during earlier stages of development. The main limitation lies in its reliance on well-initialized transition and emission matrices. Without a strong prior or carefully tuned initialization, the randomly generated starting parameters often lead to unstable convergence, poor accuracy, and high variance in results, especially for multi-trap RTS cases. This makes HMM-based digitization unreliable and unrepresentative of the true signal dynamics, especially when applied in a fully autonomous setting without manual tuning. For these reasons, it is excluded from the formal benchmark comparisons. For clarity, proximity-based, HMM-based, and Bayesian digitization strategies differ primarily in their sensitivity to residual noise, reliance on parameter initialization, and computational cost: proximity-based methods are fast but noise-sensitive, HMM-based methods are parameter-dependent and unstable without careful initialization, while the proposed Bayesian digitizer offers a lightweight and fully autonomous alternative suited for fluctuating denoised RTS signals.

Although the DTCWT denoiser is specifically optimized to suppress high-frequency white noise with excellent computational efficiency, it may not fully eliminate residual low-frequency components, such as pink noise, that often persist in RTS signals. To address this, and to further improve upon the capabilities of deep learning-based approaches like the DAEUBL, we introduce a novel Bayesian digitization method designed to complement the DTCWT. In this dual-stage architecture, the DTCWT handles the initial denoising to suppress white noise, while the Bayesian digitizer targets the remaining pink noise, enabling robust state classification across diverse RTS profiles, without neural networks or GPU-intensive processing. The proposed Bayesian digitizer applies probabilistic inference to assign each point in the denoised signal to one of the identified KDE peaks shown in \autoref{alg:bayesian_digitization}.

\begin{algorithm}[H]
\caption{Bayesian Digitization}\label{alg:bayesian_digitization}
\begin{algorithmic}[1]
\Require Denoised RTS, KDE peak levels $\mu_i$
\Ensure Digitized RTS represented by binary states of each trap
\State Estimate noise level $\sigma$ from residuals between signal and nearest KDE peak
\State Initialize uniform prior probabilities over KDE peaks
\For{each amplitude a(t) in the denoised RTS at time step t}
    \State Compute likelihoods:
    \Statex \hspace{\algorithmicindent} $L_i = \exp\left(-\frac{(a(t) - \mu_i)^2}{2\sigma^2}\right) / (\sqrt{2\pi}\sigma)$
    \State Compute posteriors: $P_i = \frac{\text{prior}_i \cdot L_i}{\sum_j \text{prior}_j \cdot L_j}$
    \State Assign current value to index $k = \arg\max_i P_i$
    \State Store $k$ in classification array
    \State Update prior: $\text{prior} \leftarrow \alpha \cdot \text{prior} + (1 - \alpha) \cdot P$
\EndFor
\State Initialize output array for binary trap state levels
\For{each classification index $k$}
    \State Convert index $k$ to binary string of length $N_{\text{trap}}$
    \State Store binary values in output array
\EndFor
\State Save digitized RTS
\end{algorithmic}
\end{algorithm}

Bayesian digitization is formulated as a probabilistic inference problem. Let $a(t)\in\mathbb{R}$ denote the denoised RTS amplitude at discrete time index $t=1,\dots,T$. From the denoised signal, we extract a finite set of $K$ discrete amplitude levels $\{\mu_k\}_{k=1}^K$, obtained as KDE peak locations. We introduce a latent discrete random variable $z_t \in \{1,\dots,K\}$, which indicates which KDE level is active at time $t$. We consider the observation model $a(t) = \mu_{z_t} + \epsilon_t$, where $\epsilon_t$ represents the marginal residual noise remaining after denoising. Under this model, digitization corresponds to inferring the most likely discrete state $z_t$ given the observed amplitude $a(t)$. We assume that the residual noise $\epsilon_t$ is approximately zero-mean Gaussian when considered marginally $p\!\left(a(t)\mid z_t=k\right) = \mathcal{N}\!\left(a(t);\mu_k,\sigma^2\right)$. This assumption implies that amplitudes close to $\mu_k$ are more likely to arise from state $k$, while deviations are penalized exponentially. The variance $\sigma^2$ controls how forgiving the classifier is with respect to deviations from KDE levels. Although RTS noise may exhibit temporal correlation (e.g., pink noise), the Gaussian assumption is applied only to the marginal sample distribution. Temporal stability is handled separately through prior smoothing.

The residual noise standard deviation $\sigma$ is estimated directly from the denoised signal by measuring deviations from the nearest KDE level. For each time step, we define $\hat{\mu}(a(t)) = \arg\min_{\mu_k} \lvert a(t) - \mu_k \rvert$, and compute the root-mean-square residual between the signal and its closest KDE level. Then Bayesian inference is performed at each time step. Let $\pi^{(t)}_k \triangleq p(z_t=k)$ denote the prior probability of state $k$ at time $t$. At initialization, we use a uniform prior $\pi^{(1)}_k = \frac{1}{K}$, reflecting equal occupancy probability across all KDE levels. Given an observation $a(t)$, the likelihood for each state $k$ is computed using a Gaussian emission model $L_k = p(a(t)\mid z_t=k)$. Applying Bayes' rule yields the posterior distribution $p(z_t=k\mid a(t))$. This update is computed in closed form and requires no iterative optimization. Lastly, a hard digitized label is obtained via maximum a posteriori (MAP) decoding $\hat{z}_t = \arg\max_k p(z_t=k\mid a(t))$. The selected index corresponds to the digitized KDE level at time $t$. To suppress spurious switching caused by residual noise and enforce temporal persistence, we update the prior distribution using exponential smoothing $\pi^{(t+1)} \leftarrow \alpha\,\pi^{(t)} + (1-\alpha)\,p(z_t\mid a(t))$, where $0<\alpha<1$. This update acts as a lightweight temporal regularizer; rather than learning a full transition matrix as in HMM, we impose a persistence bias that favors remaining in the same state unless successive observations consistently support a transition to a different dwell time in RTS. Because this update is a convex combination of valid probability distributions, it preserves normalization and non-negativity $\sum_k \pi^{(t)}_k = 1$, $\pi^{(t)}_k \ge 0$. Additionally, an optional refinement step using Dempster-Shafer Theory \cite{dempster_shafer} can be incorporated, leveraging evidence from a short window of previous classifications to further stabilize predictions, especially in multi-trap signals where amplitude levels are closely spaced.

The Bayesian digitization method is computationally efficient and scalable. Because the posterior is computed in closed form at each time step, the method avoids iterative optimization as in variational Bayes or Markov chain Monte Carlo. The computational complexity scales linearly $\mathcal{O}(LK)$ with signal length $L$ and number of KDE levels $K$. Compared to HMM-based approaches, which jointly learn emission and transition parameters via expectation-maximization and are sensitive to initialization, our method fixes emission means using data-driven KDE peaks and enforces temporal stability through prior smoothing. Prioritizing robustness, scalability, and fully autonomous analysis of long, high-resolution RTS measurements.

A final key component of RTS analysis involves characterizing the temporal dynamics of trap activity in terms of the dwell time statistics. Each RTS signal contains periods where the trap remains in either a high or low state ($\tau_{\text{high}}$ and $\tau_{\text{low}}$) as illustrated in \autoref{fig:fig1_rts_examples}a. These intervals vary in length and occur frequently, which poses a challenge to concise yet accurate summarization. Statistically, these dwell times follow a Poissonian process, leading to exponentially distributed interval lengths. We model the distribution of dwell times using the function $f(x;\bar{\tau},a) = \frac{1}{\bar{\tau}} \exp\left(-\frac{x - a}{\bar{\tau}}\right)$, where $x$ is the observed dwell time, $\bar{\tau}$ is the mean dwell time (either $\bar{\tau}_{\text{high}}$ or $\bar{\tau}_{\text{low}}$), and $a$ is an offset parameter accounting for the shift in the decay (typically negligible in RTS studies). To extract these statistics, we analyze the binary time-series outputs from the digitization stage, where each binary signal represents the activity of an individual trap over time, 1 for the high state and 0 for the low state. The extractor identifies all state transitions in the binary trace and records the durations spent in each state. These durations are compiled into histograms and fitted with the exponential decay model to compute the characteristic means, $\bar{\tau}_{\text{high}}$ and $\bar{\tau}_{\text{low}}$.

\subsection*{Four benchmarking models}

For the selection of four benchmarking models: \textbf{MA + Proximity}, \textbf{FFT + Proximity}, \textbf{DAEUBL + Proximity}, and our proposed \textbf{DTCWT + Bayesian}, we do not include every possible combination of three denoising and two digitization options due to the following reasons. With scalability constraints, benchmarking all method combinations across large-scale datasets in terabytes is computationally infeasible. Secondly, empirical redundancy, preliminary tests of cross-method combinations (e.g., DAEUBL + Bayesian digitization) showed no meaningful quality gains over proximity-based digitization in those contexts. In fact, using Bayesian digitization after DAEUBL was not beneficial. While DAEUBL produces high-confidence outputs even when RTS levels are biased or distorted, the Bayesian digitizer was found less compatible; its design assumes uncertainty structures typical of DTCWT outputs, which tend to fluctuate around RTS levels. On the other hand, DTCWT + proximity digitization was also not pursued, as proximity-based digitizers inherently lack the capability to handle pink noise, a limitation they share with baseline combinations like \textbf{MA + Proximity} and \textbf{FFT + Proximity}.

For benchmarking fairness, all methods evaluated in this study are blind to the synthetic benchmark data: none of the denoisers or digitizers are optimized, trained, or tuned on the evaluated RTS signals. The adaptive rules used in the proposed DTCWT denoiser are derived independently from a separate set of real RTS measurements and rely only on general signal statistics, as detailed in \textbf{Denoising and KDE Algorithm}. Baseline methods such as MA and FFT employ adaptive parameters consistent with their standard usage in RTS analysis, while DAEUBL is evaluated using its available pretrained model without retraining. To ensure fair comparison, we adopt a unified and more aggressive noise-scaling scheme across all methods, avoiding the use of the original DAEUBL training data, which would otherwise introduce an unfair advantage.

\section*{Data availability}

The datasets used and/or analysed during the current study are available from the corresponding author on reasonable request.

\section*{Acknowledgements}

We acknowledge HeeBong Yang for providing carbon nanotube film RTS data and the Digital Research Alliance of Canada for computing resources for our calculations.

\section*{Funding}

This work was supported by Industry Canada, the Ontario Ministry of Research \& Innovation through Early Researcher Awards (RE09-068), and the Canada First Research Excellence Fund-Transformative Quantum Technologies (CFREF-TQT).

\section*{Author contributions}

N.Y.K. conceived the research idea and supervised the whole study. T.B. investigated the data synthesis, denoising, and digitization aspects of the analysis. T.B. wrote the code for synthesizing the datasets, DTCWT denoising, KDE steps, and data analysis. A.K. wrote the code for Bayesian digitization. T.B. performed simulations and collected simulation error statistics. T.B. prepared all figures. T.B. and N.Y.K. wrote the paper.

\bibliographystyle{IEEEtran}
\bibliography{citations}

@article{rts.SR,
    author = "Marcel Robitaille and HeeBong Yang and Lu Wang and Bowen Deng and Na Young Kim",
    title = "Deep neural network analysis models for complex random telegraph signals",
    journal = "Scientific Reports",
    year = "2024",
    publisher = "Nature",
    address = "200 University Ave W, Waterloo, ON N2L 3G1, Canada",
    note = "Institute for Quantum Computing, Department of Electrical and Computer Engineering, Waterloo Institute for Nanotechnology, Department of Chemistry, and Perimeter Institute, University of Waterloo"
}

@article{bowen.thesis,
    title = {A denoising autoencoder based on U-Net and bidirectional long short-term memory for multi-level random telegraph signal analysis},
    journal = {Engineering Applications of Artificial Intelligence},
    volume = {135},
    pages = {108685},
    year = {2024},
    issn = {0952-1976},
    doi = {https://doi.org/10.1016/j.engappai.2024.108685},
    url = {https://www.sciencedirect.com/science/article/pii/S0952197624008431},
    author = {Bowen Deng and HeeBong Yang and Na Young Kim}
}

@INPROCEEDINGS{wavelet_methods,
  author={Mortazavi, S. H. and Shahrtash, S. M.},
  booktitle={2008 43rd International Universities Power Engineering Conference}, 
  title={Comparing denoising performance of DWT,WPT, SWT and DT-CWT for Partial Discharge signals}, 
  year={2008},
  volume={},
  number={},
  pages={1-6},
  doi={10.1109/UPEC.2008.4651625}}

@misc{dtcwt,
  author = {Nick Kingsbury},
  title = {Dual-Tree Complex Wavelets: Part 1},
  howpublished = {Talk presentation, PDF format},
  institution = {Signal Processing Group, Dept. of Engineering, University of Cambridge},
  year = {2005},
  month = {February},
  url = {http://www.eng.cam.ac.uk/~ngk},
  note = {Accessed: 2024-05},
}

@article{RTS_studies1,
title = {Analysis of random telegraph noise in resistive memories: The case of unstable filaments},
journal = {Micro and Nano Engineering},
volume = {19},
pages = {100205},
year = {2023},
issn = {2590-0072},
doi = {https://doi.org/10.1016/j.mne.2023.100205},
url = {https://www.sciencedirect.com/science/article/pii/S2590007223000357},
author = {Nikolaos Vasileiadis and Alexandros Mavropoulis and Panagiotis Loukas and Georgios Ch. Sirakoulis and Panagiotis Dimitrakis}
}

@article{RTS_studies2,
title = {Enhanced statistical detection of random telegraph noise in frequency and time domain},
journal = {Solid-State Electronics},
volume = {194},
pages = {108320},
year = {2022},
issn = {0038-1101},
doi = {https://doi.org/10.1016/j.sse.2022.108320},
url = {https://www.sciencedirect.com/science/article/pii/S0038110122000922},
author = {Owen Gauthier and Sébastien Haendler and Patrick Scheer and Alexandre Vernhet and Quentin Rafhay and Christoforos Theodorou}
}

@techreport{dempster_shafer,
  title        = {Combination of Evidence in Dempster-Shafer Theory},
  author       = {Sentz, Karl and Ferson, Scott},
  year         = {2002},
  number       = {SAND2002-0835},
  institution  = {Sandia National Laboratories},
  address      = {Albuquerque, New Mexico and Livermore, California},
  month        = {April},
  note         = {Approved for public release; further dissemination unlimited. Prepared under Contract DE-AC04-94AL85000.},
}

@mastersthesis{jackie.thesis,
    author = "Tonghe Bai",
    title = "Towards a Robust Framework for Analyzing Random Telegraph Signals (RTS): Application to 2-level RTS in a Semiconductor Quantum Dot",
    year = "2024",
    school = "University of Waterloo",
    address = "200 University Ave. West, Waterloo, N2L 3G1, Ontario, Canada",
    note = "Department of Electrical and Computer Engineering, Institute for Quantum Computing"
}

@article{qd.SR,
    author = "J. Kerski and H. Mannel and P. Lochner and E. Kleinherbers and A. Kurzmann and A. Ludwig and A. D. Wieck and J. König and A. Lorke and M. Geller",
    title = "Post-processing of real-time quantum event measurements for an optimal bandwidth",
    journal = "Scientific Reports",
    year = "2023",
    volume = "13",
    pages = "1105",
    publisher = "Nature",
    address = "",
    note = "Faculty of Physics and CENIDE, University of Duisburg-Essen. 2nd Institute of Physics, RWTH Aachen University. Chair for Applied Solid State Physics, Ruhr-Universität Bochum"
}

@article{mosfets,
title = {Model for random telegraph signals in sub-micron MOSFETS},
journal = {Solid-State Electronics},
volume = {47},
number = {9},
pages = {1443-1449},
year = {2003},
issn = {0038-1101},
doi = {https://doi.org/10.1016/S0038-1101(03)00100-X},
url = {https://www.sciencedirect.com/science/article/pii/S003811010300100X},
author = {Nuditha Vibhavie Amarasinghe and Zeynep Çelik-Butler and Anna Zlotnicka and Fang Wang}
}

@article{transistor,
  title={Anomalous random telegraph noise in nanoscale transistors as direct evidence of two metastable states of oxide traps},
  author={Guo, S and Wang, R and Mao, D et al.},
  journal={Scientific Reports},
  volume={7},
  number={1},
  pages={6239},
  year={2017},
  publisher={Nature Publishing Group},
  doi={10.1038/s41598-017-06467-7}
}

@misc{apd,
  author = {Paschotta, R.},
  publisher = {RP Photonics AG},
  title = {Avalanche Photodiodes},
  year = {},
  month = {},
  howpublished = {RP Photonics Encyclopedia},
  note = {{Available online at \url{https://www.rp-hotonics.com/avalanche_photodiodes.html}}},
  url = {https://www.rp-photonics.com/avalanche_photodiodes.html},
  doi = {10.61835/pbn},
  urldate = {2024-10-18}
}

@Inbook{low_freq,
author="Fleetwood, D. M.",
title="Origins of 1/f Noise in Electronic Materials and Devices: A Historical Perspective",
bookTitle="Noise in Nanoscale Semiconductor Devices",
year="2020",
publisher="Springer International Publishing",
address="Cham",
pages="1--31",
isbn="978-3-030-37500-3",
doi="10.1007/978-3-030-37500-3_1",
url="https://doi.org/10.1007/978-3-030-37500-3_1"
}

@article{rts_noise,
doi = {10.1088/1742-5468/ac5dbf},
url = {https://dx.doi.org/10.1088/1742-5468/ac5dbf},
year = {2022},
month = {apr},
publisher = {IOP Publishing and SISSA},
volume = {2022},
number = {4},
pages = {043201},
author = {Roberto da Silva and Gilson Wirth},
title = {{RTS} noise in semiconductor devices: time constants estimates and observation window analysis},
journal = {Journal of Statistical Mechanics: Theory and Experiment}
}

@ARTICLE{rts.SR.ref1,
  author={Hung, K.K. and Ko, P.K. and Hu, C. and Cheng, Y.C.},
  journal={IEEE Electron Device Letters}, 
  title={Random telegraph noise of deep-submicrometer MOSFETs}, 
  year={1990},
  volume={11},
  number={2},
  pages={90-92},
  doi={10.1109/55.46938},
  ISSN={1558-0563},
  month={Feb}}

@INPROCEEDINGS{rts.SR.ref2,
  author={Wang, Xinyang and Rao, Padmakumar R. and Mierop, Adri and Theuwissen, Albert J.P.},
  booktitle={2006 International Electron Devices Meeting}, 
  title={Random Telegraph Signal in {CMOS} Image Sensor Pixels}, 
  year={2006},
  volume={},
  number={},
  pages={1-4},
  doi={10.1109/IEDM.2006.346973},
  ISSN={2156-017X},
  month={Dec}}

@article{rts.SR.ref3,
  title = {Kinetics of nonequilibrium quasiparticle tunneling in superconducting charge qubits},
  author = {Shaw, M. D. and Lutchyn, R. M. and Delsing, P. and Echternach, P. M.},
  journal = {Phys. Rev. B},
  volume = {78},
  issue = {2},
  pages = {024503},
  numpages = {9},
  year = {2008},
  month = {Jul},
  publisher = {American Physical Society},
  doi = {10.1103/PhysRevB.78.024503},
  url = {https://link.aps.org/doi/10.1103/PhysRevB.78.024503}
}

@article{rts.SR.ref5,
  title = {How to enhance dephasing time in superconducting qubits},
  author = {Cywi\ifmmode \acute{n}\else \'{n}\fi{}ski, \L{}ukasz and Lutchyn, Roman M. and Nave, Cody P. and Das Sarma, S.},
  journal = {Phys. Rev. B},
  volume = {77},
  issue = {17},
  pages = {174509},
  numpages = {11},
  year = {2008},
  month = {May},
  publisher = {American Physical Society},
  doi = {10.1103/PhysRevB.77.174509},
  url = {https://link.aps.org/doi/10.1103/PhysRevB.77.174509}
}

@Book{hmm.language,
  author =       "Daniel Jurafsky and James H. Martin",
  title =        "Speech and Language Processing: An Introduction to
                 Natural Language Processing, Computational Linguistics,
                 and Speech Recognition with Language Models",
  year =         "2024",
  publisher = "",
  url = {https://web.stanford.edu/~jurafsky/slp3/},
  note = "Online manuscript released August 20, 2024",
  edition =         "3rd"
}

@inproceedings{hmm.harvard,
  title={Introduction to Hidden Markov Models},
  booktitle="",
  author={Alperen Değirmenci},
  year={2015},
  url={https://api.semanticscholar.org/CorpusID:45511616}
}

@article{hmm.nature,
  author = {Sean R. Eddy},
  title = {What is a hidden Markov model?},
  journal = {Nature Biotechnology},
  year = {2004},
  volume = {22},
  number = {10},
  pages = {1315--1316},
  doi = {10.1038/nbt1004-1315},
  url = {https://doi.org/10.1038/nbt1004-1315},
  issn = {1546-1696}
}

@article{rtn.bayesian,
  title={Bayesian Estimation of Multi-Trap RTN Parameters Using Markov Chain Monte Carlo Method},
  author={Hiromitsu AWANO and Hiroshi TSUTSUI and Hiroyuki OCHI and Takashi SATO},
  journal={IEICE Transactions on Fundamentals of Electronics, Communications and Computer Sciences},
  volume={E95.A},
  number={12},
  pages={2272-2283},
  year={2012},
  doi={10.1587/transfun.E95.A.2272}
}

@article{rtn.bayesian2,
author = {Chen, Xiaoming and Wang, Lin and Li, Boxun and Wang, Yu and Li, Xin and Liu, Yongpan and Yang, Huazhong},
year = {2015},
month = {01},
pages = {1-1},
title = {Modeling Random Telegraph Noise as a Randomness Source and Its Application in True Random Number Generation},
volume = {35},
journal = {IEEE Transactions on Computer-Aided Design of Integrated Circuits and Systems},
doi = {10.1109/TCAD.2015.2511074}
}

@article{biochem,
title = {Fluctuation-driven directional flow in biochemical cycle: further study of electric activation of Na,K pumps},
journal = {Biophysical Journal},
volume = {72},
number = {6},
pages = {2496-2502},
year = {1997},
issn = {0006-3495},
doi = {https://doi.org/10.1016/S0006-3495(97)78894-5},
url = {https://www.sciencedirect.com/science/article/pii/S0006349597788945},
author = {T.D. Xie and Y. Chen and P. Marszalek and T.Y. Tsong}
}

@article{gene,
author = {Hansen, Maike and Desai, Ravi and Simpson, Mike and Weinberger, Leor},
year = {2018},
month = {09},
pages = {},
title = {Cytoplasmic Amplification of Transcriptional Noise Generates Substantial Cell-to-Cell Variability},
volume = {7},
journal = {Cell Systems},
doi = {10.1016/j.cels.2018.08.002}
}

@INPROCEEDINGS{sensors,
  author={Wang, Xinyang and Rao, Padmakumar R. and Mierop, Adri and Theuwissen, Albert J.P.},
  booktitle={2006 International Electron Devices Meeting}, 
  title={Random Telegraph Signal in CMOS Image Sensor Pixels}, 
  year={2006},
  volume={},
  number={},
  pages={1-4},
  doi={10.1109/IEDM.2006.346973}}

@article{MA.reram,
title = {Analysis of random telegraph noise in resistive memories: The case of unstable filaments},
journal = {Micro and Nano Engineering},
volume = {19},
pages = {100205},
year = {2023},
issn = {2590-0072},
doi = {https://doi.org/10.1016/j.mne.2023.100205},
url = {https://www.sciencedirect.com/science/article/pii/S2590007223000357},
author = {Nikolaos Vasileiadis and Alexandros Mavropoulis and Panagiotis Loukas and Georgios Ch. Sirakoulis and Panagiotis Dimitrakis}
}

@ARTICLE{fft,
  author={Brigham, E. O. and Morrow, R. E.},
  journal={IEEE Spectrum}, 
  title={The fast Fourier transform}, 
  year={1967},
  volume={4},
  number={12},
  pages={63-70},
  doi={10.1109/MSPEC.1967.5217220}}

@article{emd,
author = {Moshrefi, Amirhossein and Aghababa, Hossein and Shoaei, Omid},
year = {2021},
month = {01},
pages = {90-96},
title = {Employing the Empirical Mode Decomposition to Denoise the Random Telegraph Noise},
volume = {34},
journal = {Engineering Transactions},
doi = {10.5829/ije.2021.34.01a.11}
}

@article{rts_ml_RNN,
  title = {Random telegraph signal analysis with a recurrent neural network},
  author = {Lambert, N. J. and Esmail, A. A. and Edwards, M. and Ferguson, A. J. and Schwefel, H. G. L.},
  journal = {Phys. Rev. E},
  volume = {102},
  issue = {1},
  pages = {012312},
  numpages = {5},
  year = {2020},
  month = {Jul},
  publisher = {American Physical Society},
  doi = {10.1103/PhysRevE.102.012312},
  url = {https://link.aps.org/doi/10.1103/PhysRevE.102.012312}
}

@misc{rts_ml_RTNinja,
      title={RTNinja: a generalized machine learning framework for analyzing random telegraph noise signals in nanoelectronic devices}, 
      author={Anirudh Varanasi and Robin Degraeve and Philippe Roussel and Clement Merckling},
      year={2025},
      eprint={2507.08424},
      archivePrefix={arXiv},
      primaryClass={cs.LG},
      url={https://arxiv.org/abs/2507.08424}, 
}

@article{rts_ml_SOM,
author = {González-Cordero, Gerardo and Gonzalez, Mireia and Morell, A and Jiménez-Molinos, F. and Campabadal, Francesca and Roldan, Juan},
year = {2020},
month = {01},
pages = {},
title = {Neural network based analysis of random telegraph noise in resistive random access memories},
volume = {35},
journal = {Semiconductor Science and Technology},
doi = {10.1088/1361-6641/ab6103}
}

@INPROCEEDINGS{rts_ml_clustering,
  author={Li, Xinze and Sun, Ying and Wan, Jing and Chen, Bing and Cheng, Ran and Han, Genquan},
  booktitle={2022 International Conference on IC Design and Technology (ICICDT)}, 
  title={Machine Learning Method for Accurate Analysis of Complicated Low Temperature Random Telegraph Noise}, 
  year={2022},
  volume={},
  number={},
  pages={20-23},
  doi={10.1109/ICICDT56182.2022.9933107}}

@misc{mcmc1,
title={MCMC-Based Inference in the Era of Big Data: A Fundamental Analysis of the Convergence Complexity of High-Dimensional Chains}, 
author={Bala Rajaratnam and Doug Sparks},
year={2015},
eprint={1508.00947},
archivePrefix={arXiv},
primaryClass={math.ST},
url={https://arxiv.org/abs/1508.00947}, 
}

@article{mcmc2,
author = {Craiu, Radu V. and Gustafson, Paul and Rosenthal, Jeffrey S.},
title = {Reflections on Bayesian inference and Markov chain Monte Carlo},
journal = {Canadian Journal of Statistics},
volume = {50},
number = {4},
pages = {1213-1227},
doi = {https://doi.org/10.1002/cjs.11707},
url = {https://onlinelibrary.wiley.com/doi/abs/10.1002/cjs.11707},
eprint = {https://onlinelibrary.wiley.com/doi/pdf/10.1002/cjs.11707},
year = {2022}
}

@article{rts_wavelet,
title = {1/f Noise decomposition in random telegraph signals using the wavelet transform},
journal = {Physica A: Statistical Mechanics and its Applications},
volume = {380},
pages = {75-97},
year = {2007},
issn = {0378-4371},
doi = {https://doi.org/10.1016/j.physa.2007.02.111},
url = {https://www.sciencedirect.com/science/article/pii/S0378437107002026},
author = {Fabio Principato and Gaetano Ferrante}
}

@article{Alessandro_Romeo_1,
    author = {Romeo, Alessandro B. and Horellou, Cathy and Bergh, Jöran},
    title = {N-body simulations with two-orders-of-magnitude higher performance using wavelets},
    journal = {Monthly Notices of the Royal Astronomical Society},
    volume = {342},
    number = {2},
    pages = {337-344},
    year = {2003},
    month = {06},
    issn = {0035-8711},
    doi = {10.1046/j.1365-8711.2003.06549.x},
    url = {https://doi.org/10.1046/j.1365-8711.2003.06549.x},
    eprint = {https://academic.oup.com/mnras/article-pdf/342/2/337/3404426/342-2-337.pdf},
}

@article{Alessandro_Romeo_2,
    author = {Romeo, Alessandro B. and Horellou, Cathy and Bergh, Jöran},
    title = {A wavelet add‐on code for new‐generation N‐body simulations and data de‐noising (JOFILUREN)},
    journal = {Monthly Notices of the Royal Astronomical Society},
    volume = {354},
    number = {4},
    pages = {1208-1222},
    year = {2004},
    month = {11},
    issn = {0035-8711},
    doi = {10.1111/j.1365-2966.2004.08303.x},
    url = {https://doi.org/10.1111/j.1365-2966.2004.08303.x},
    eprint = {https://academic.oup.com/mnras/article-pdf/354/4/1208/3597122/354-4-1208.pdf},
}

@article{Alessandro_Romeo_3,
doi = {10.1086/591236},
url = {https://doi.org/10.1086/591236},
year = {2008},
month = {oct},
publisher = {},
volume = {686},
number = {1},
pages = {1},
author = {Romeo, Alessandro B. and Agertz, Oscar and Moore, Ben and Stadel, Joachim},
title = {Discreteness Effects in {$\Lambda$}CDM Simulations: A Wavelet-Statistical View},
journal = {The Astrophysical Journal}
}

\end{document}